\documentclass[english]{sobraep}

\usepackage[
    backend=biber,
    style=nature,
  ]{biblatex}
\usepackage{csquotes}

\title{On Classifying Images using Quantum Image Representation}

\author{Ankit Khandelwal$^{1,2}$, M Girish Chandra$^{2}$, Sayantan Pramanik$^{3}$\\
	\normalsize $^{1}$Centre for High Energy Physics, Indian Institute of Science, Bengaluru, India \\
	\normalsize $^{2}$TCS Research, India\\
	\normalsize $^{3}$TCS Incubation, India\\
	\normalsize e-mail: \href{mailto:ankit27.kh@gmail.com}{ankit27.kh@gmail.com}, \href{mailto:m.gchandra@tcs.com}{m.gchandra@tcs.com}, \href{mailto:sayantan.pramanik@tcs.com}{sayantan.pramanik@tcs.com}
}

\begin{document}

\maketitle

\begin{abstract}
In this paper, we consider different Quantum Image Representation Methods to encode images into quantum states and then use a Quantum Machine Learning pipeline to classify the images. We provide encouraging results on classifying benchmark datasets of grayscale and colour images using two different classifiers. We also test multi-class classification performance.
\end{abstract}


\section{INTRODUCTION}

Quantum Image Representation (QIR) is a catch-all term for methods to encode an image as a quantum state. General data encoding methods \cite{PhysRevA.102.032420} exist, such as Angle Encoding and Amplitude Encoding. But these do not take advantage of the specific structure of image data. Thus, unique representation methods have been invented to tackle image data and its representation as a quantum state using a quantum circuit.

A number of methods can be found in the literature. We have considered Flexible Representation of Quantum Images (FRQI) \cite{Le2011} for grayscale images and Multi-Channel Representation for Quantum Image (MCQI) \cite{Sun2013AnRM} for colour images to encode the image and later use a quantum classifier on the encoded image to perform binary and multi-class classification.

We use a variational quantum classifier (VQC) and an autoencoder classifier (AC) to classify the images. We have obtained encouraging results using some classic benchmark datasets.

The paper is organised as follows. In Section \ref{qir}~\;, we give a brief summary of the QIR methods used in this study. In Section \ref{classifiers}~\;, we describe the quantum classifiers used for classifying the images. Section \ref{data}~~~ details the datasets used in the paper. In Section \ref{implementation}~~~ we give the implementation details. Section \ref{results}~~ presents the classification results obtained from our simulations. In Section \ref{conclude}~~~ we conclude and provide a few markers for future work.

\section{Quantum Image Representation Methods}
\label{qir}

In this section we give a summary of the QIR methods used in the text. First some details of images:
\begin{enumerate}
    \item Image dimension $=2^n\times2^n$ ($n=1$ means a $2\times2$ image $=4$ pixels)
    \item Gray scale images $\rightarrow$ Pixel values $\in[0,255] \rightarrow$ In binary $\in\{0,1\}^8$\\
        Number of Matrix Elements $=2^n\times2^n$
    \item Color images $\rightarrow$ RGB, pixel values of each channel $\in[0,255]$\\
        Number of Matrix Elements $=2^n\times2^n\times3$
\end{enumerate}

\subsection{FRQI}

FRQI encodes the image data into a quantum state given by:
\begin{equation}
    \left|I(\theta)\right>=\frac{1}{2^n}\sum_{i=0}^{2^{2n}-1}[\cos(\theta_i)\left|0\right>+\sin(\theta_i)\left|1\right>]\otimes\left|i\right>
    \label{eq_frqi}
\end{equation}
\begin{equation*}
    \theta_i\in\left[0,\frac{\pi}{2}\right], \theta=(\theta_0,\theta_1,\dots,\theta_{2^{2n}-1})
\end{equation*}
Here, $\left|0\right>$ and $\left|1\right>$ are single qubit computational basis states and $\left|i\right>, i=0,1,\dots,2^{2n}-1$ are $2n$ qubit computational basis states. The $\cos(\theta_i)\left|0\right>+\sin(\theta_i)\left|1\right>$ part is used to encode the pixel values while $\left|i\right>$ encodes the pixel location.

The circuit to encode the image can be constructed using Hadamard ($H$) and Control Rotation, $C^{2n}R_y(2\theta)$, gates. It needs to be measured multiple times to get back the image from the state. The image retrieval process is probabilistic, and the result will depend on the number of shots used.

The number of qubits used in this representation is $2n+1$ with $2n$ qubits to encode the pixel location and 1 qubit to encode the pixel values. Pixel values are encoded as angles and are thus scaled to fit in the range $\left[0,\frac{\pi}{2}\right]$.

\subsection{MCQI}
MCQI representation uses $2n+3$ qubits to encode colour images while using $2n$ qubits to encode the pixel location like FRQI and the 3 remaining qubits to encode the pixel values of the RGB channels. This encoding is inspired by FRQI.\\
MCQI encodes the image into a quantum state given by:
\begin{equation}
    \left|I(\theta)\right>=\frac{1}{2^n+1}\sum_{i=0}^{2^{2n}-1}\left|C^i_{RGB}\right>\otimes\left|i\right>
    \label{eq:mcqi}
\end{equation}
The color information is encoded in:
\begin{gather*}
   \left|C^i_{RGB}\right> =\cos(\theta^i_R)\left|000\right>+\cos(\theta^i_G)\left|001\right>+\cos(\theta^i_B)\left|010\right>\\
    +\sin(\theta^i_R)\left|100\right>+\sin(\theta^i_G)\left|101\right>+\sin(\theta^i_B)\left|110\right>\\
    +\cos(0)\left|011\right>+\sin(0)\left|111\right>
\end{gather*}

Colour encoding angle is applied to the R channel qubit using Control Rotation ($C^{2}R_y(2\theta)$) gates where it is controlled by the G and B channel qubits, and  for each pixel, 3 $C^{2n}(C^{2}R_y(2\theta))$ gates are applied to encode the position and value information. The $\theta$ values are calculated from pixel values:
\begin{equation*}
    \theta=\cos^{-1}{p}
\end{equation*}
where, $p$ is the pixel values $\in[0,1]$. We get in this range by dividing the integer pixel values $\in[0,255]$ by 255.

As before, the image retrieval process is probabilistic and depends on the number of shots.

\section{Quantum Classifiers}
\label{classifiers}

We have used two different methods to classify the images in this paper. We describe these methods below.

\subsection{Variational Quantum Classifier}

To classify the images, we need some value to distinguish the two classes. The Z expectation value $(ez)$ of the first qubit gives a natural split. We apply a variational ansatz on the quantum image and measure the expectation value of the colour qubit for FRQI and the R channel qubit for MCQI. The expectation value lies in the range $[-1,1]$, and thus a split can be formed such that:
\begin{align}
    \textbf{class} = \left\{\begin{array}{ l l }
    -1     &\textbf{if} \quad ez\leq s \\
    1    &\textbf{if} \quad ez> s
    \end{array}\right.
\end{align}
where $s$ is the split set to 0 by default but can be trained to get optimal results.

\subsubsection{Ansatz}

\begin{figure}[htbp]
    \centering
    \includegraphics[width=\columnwidth]{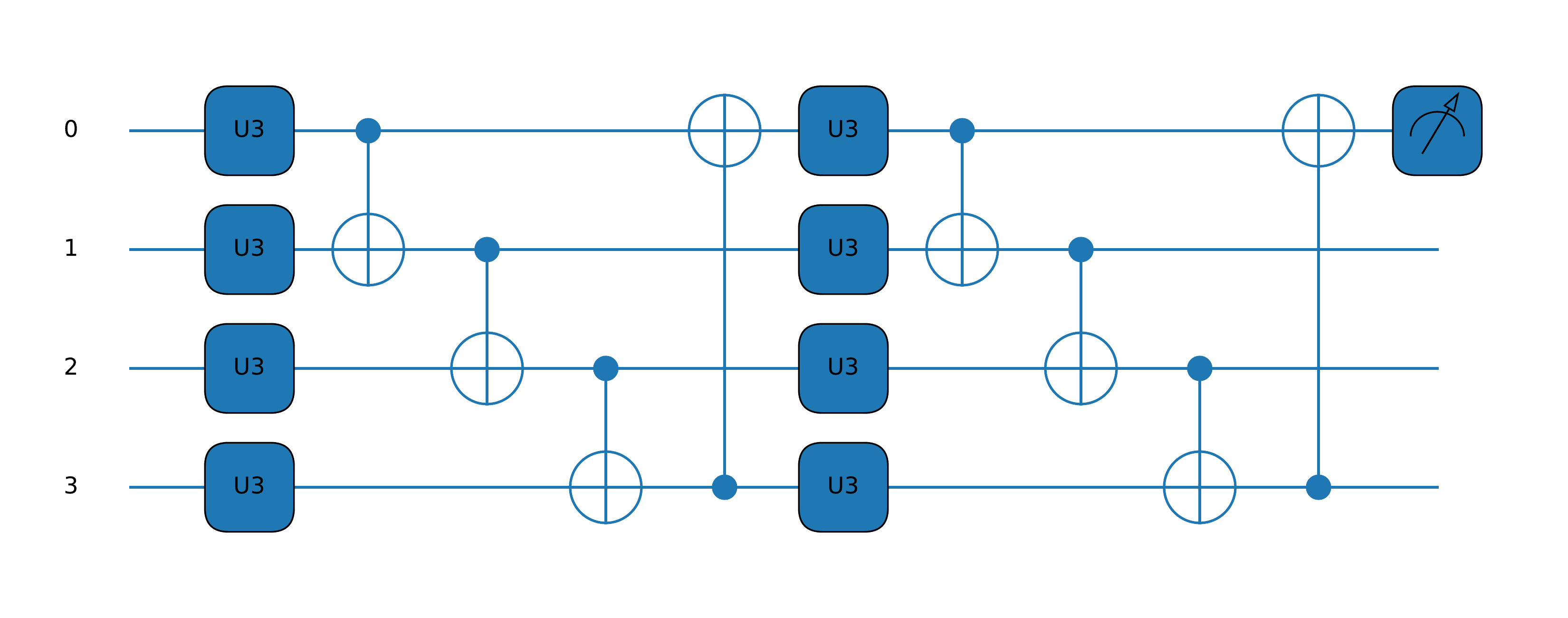}
    \caption{Variational Classifier Ansatz with 4 qubits and 2 layers. U3 denotes a single qubit general rotation gate.}
    \label{fig:vari}
\end{figure}

We use a straightforward ansatz that consists of a general single-qubit rotation on each qubit and then a layer of CNOT gates, as shown in Fig \ref{fig:vari}. The number of layers of the ansatz is a hyperparameter. The number of parameters in the classifier $=3\times N \times l$ where $N=$ number of qubits and $l=$ number of layers.

\subsection{Autoencoder Classifier}
\begin{figure}[htb]
    \centering
    \includegraphics[width=\columnwidth]{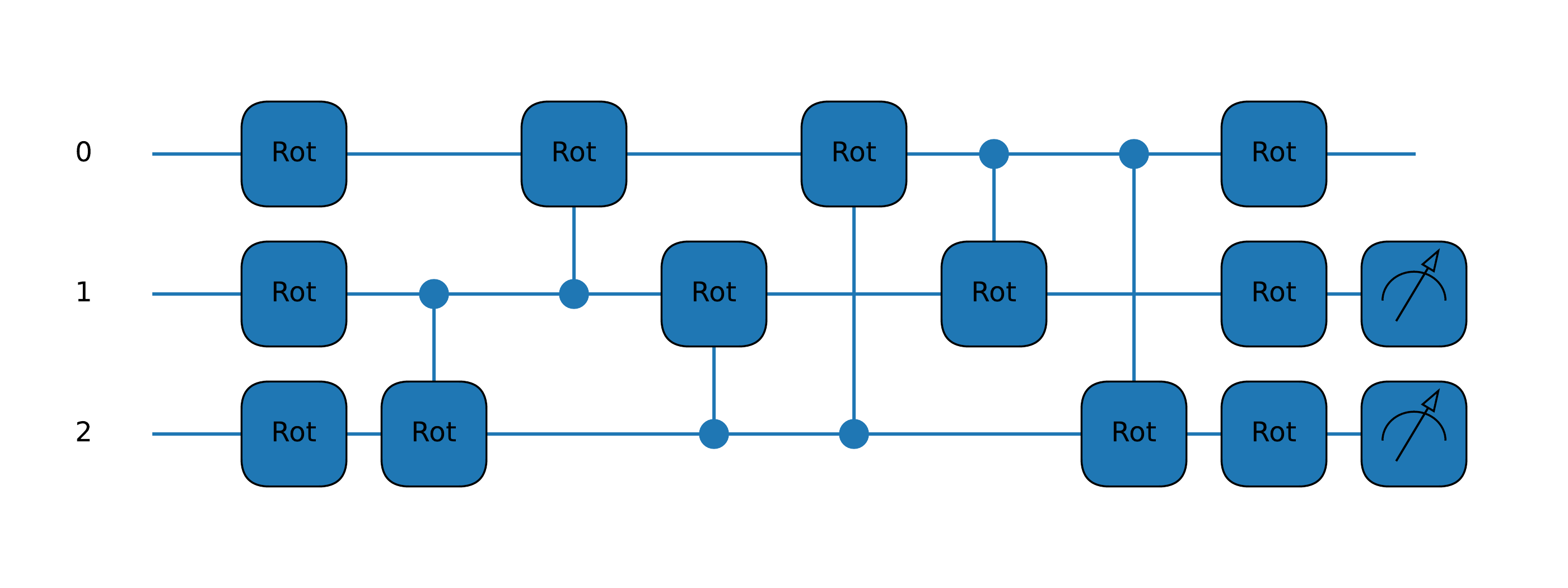}
    \caption{Autoencoder with 1 data and 2 trash qubits. 'Rots' are general single-qubit rotation gates.}
    \label{fig:auto}
\end{figure}

An autoencoder is a tool to reduce the dimension of data. We use the autoencoder's quantum analogue \cite{romero2017quantum} to compress the image state into a single qubit state. To use it as a classifier, we train the autoencoder to only compress the positive class. The autoencoder is trained by maximising the fidelity of the trash qubits with the zero state (${\left|0\right>}^{\otimes T}$) where $T$ is the number of trash qubits. Once the autoencoder is trained on the positive class of the training data, we then use it to compress the validation data. We then measure the fidelity of the trash qubits again with the zero state and classify them depending on the resulting fidelity. The positive class should have higher fidelity values than the negative class. A single layer of the autoencoder is shown in Fig \ref{fig:auto}.

\section{Dataset Details}
\label{data}
We use two grayscale and one colour image datasets.

\subsection{Bars and Stripes (BAS)}

\begin{figure}
    \centering
    \begin{subfigure}{0.45\columnwidth}
    \centering
    \includegraphics[width=\textwidth]{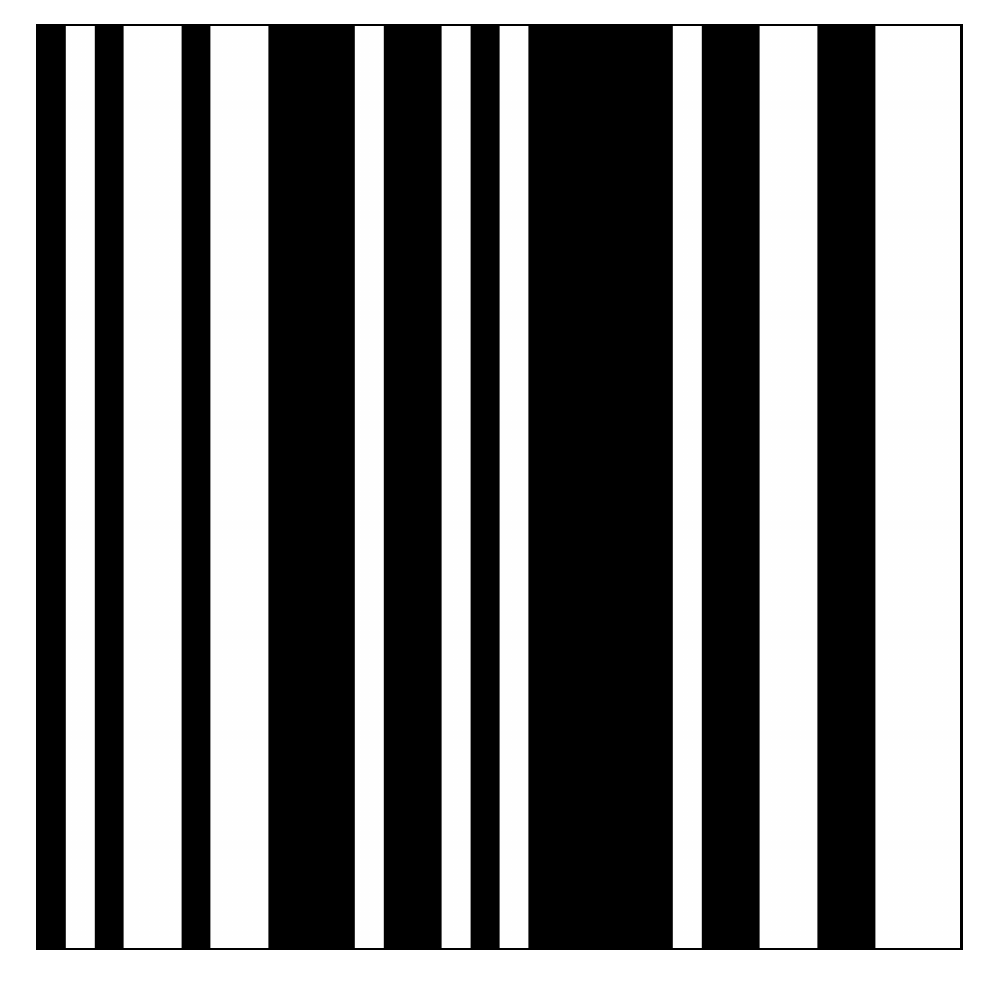}
    \end{subfigure}
    \begin{subfigure}{0.45\columnwidth}
    \centering
    \includegraphics[width=\textwidth]{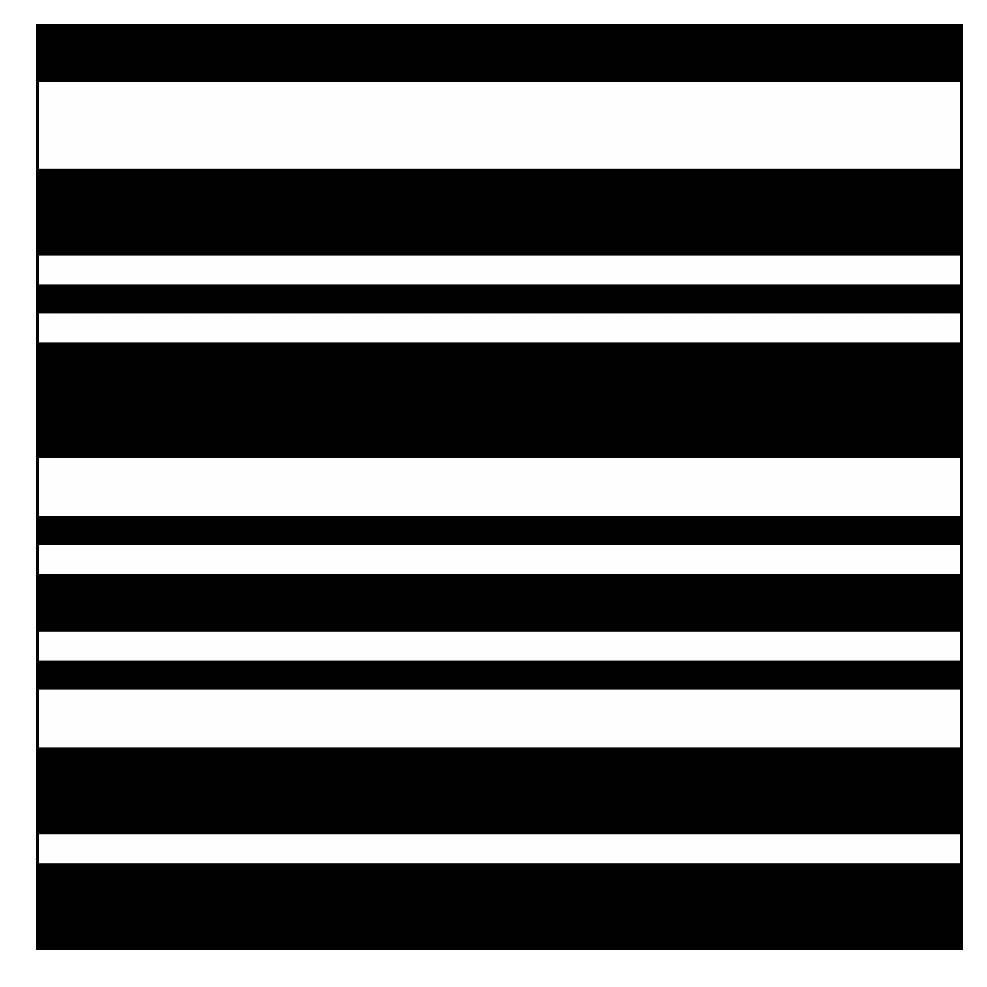}
    \end{subfigure}
    \caption{Example of the two classes in the BAS data with $n=5$.}
    \label{fig:bas}
\end{figure}

This dataset contains black and white images of dimension $2^n\times2^n$. Example images are shown in Fig \ref{fig:bas} for $n=5$. These images are randomly generated. Horizontal stripes are one class, and vertical bars are another class. This dataset is used for binary classification.

\subsection{MNIST}
\begin{figure*}
    \centering
    \begin{subfigure}{0.19\textwidth}
         \centering
         \includegraphics[width=\textwidth]{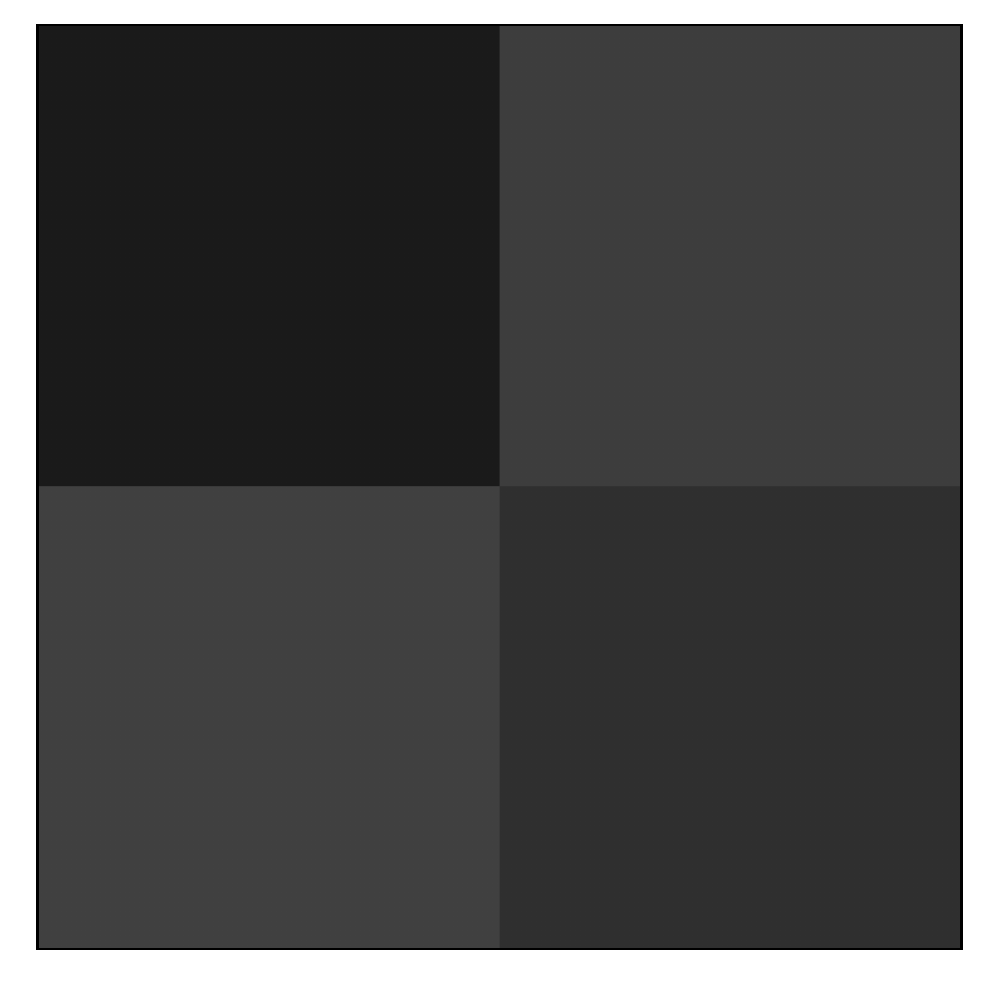}
    \end{subfigure}
    \begin{subfigure}{0.19\textwidth}
        \centering
        \includegraphics[width=\textwidth]{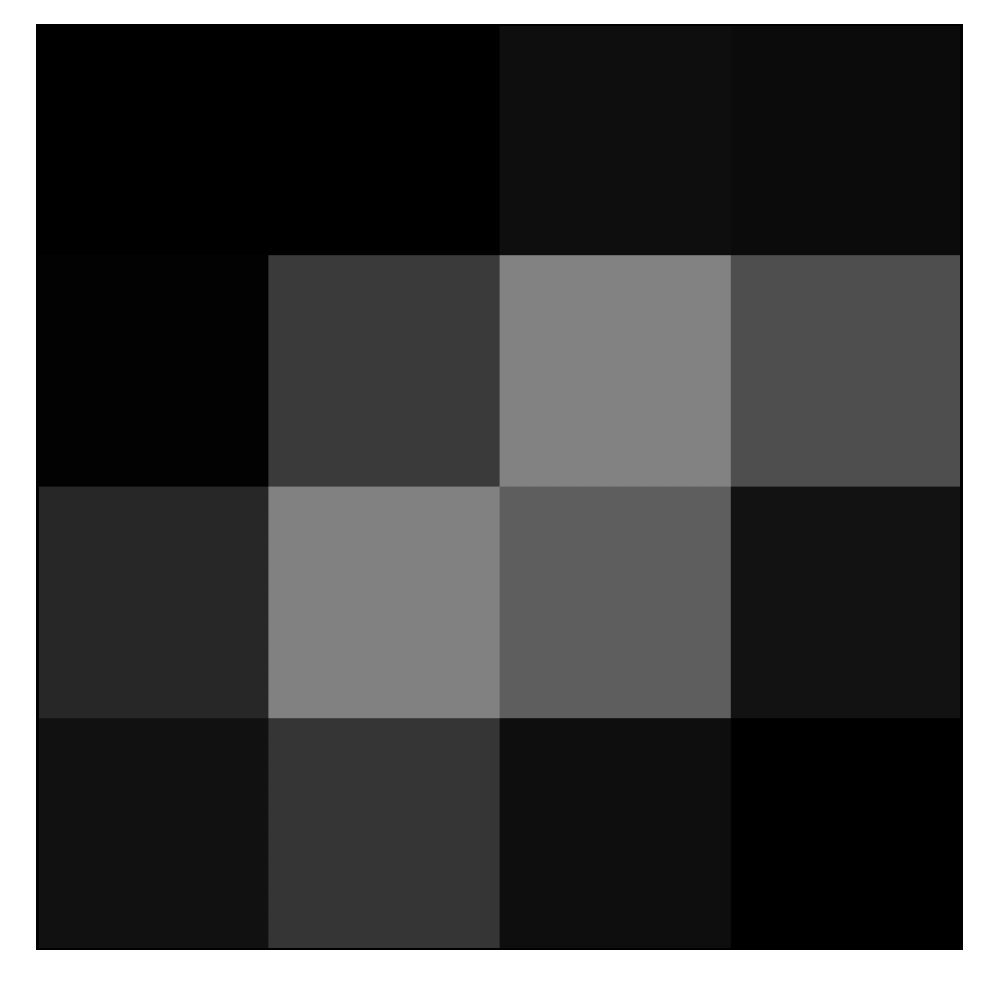}
    \end{subfigure}
    \begin{subfigure}{0.19\textwidth}
         \centering
         \includegraphics[width=\textwidth]{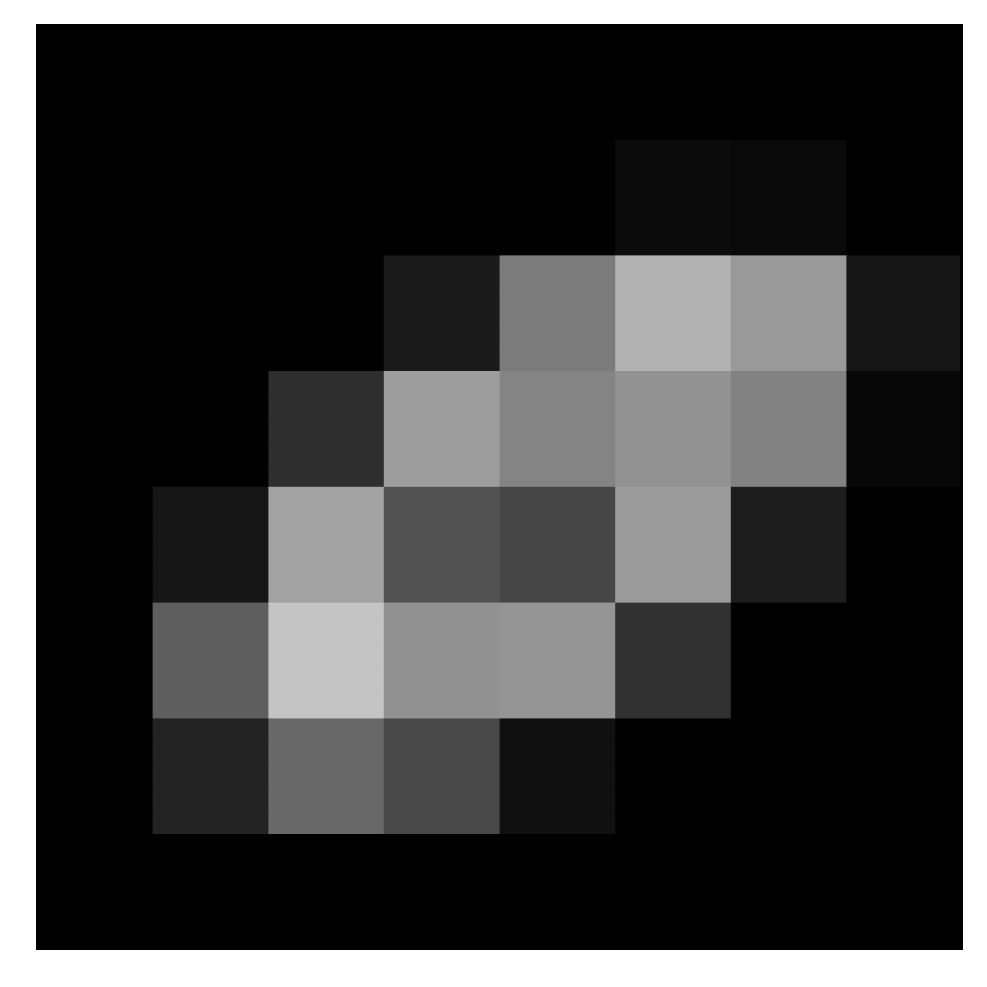}
    \end{subfigure}
    \begin{subfigure}{0.19\textwidth}
         \centering
         \includegraphics[width=\textwidth]{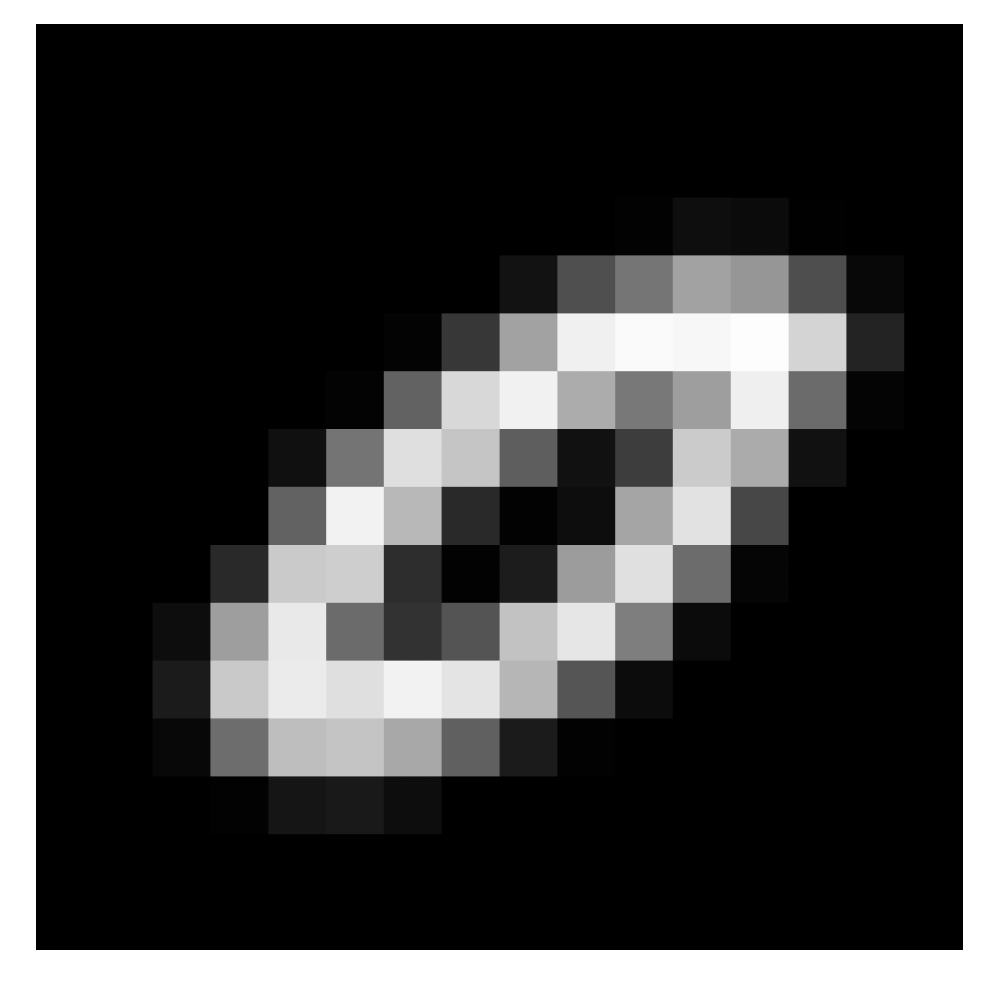}
    \end{subfigure}
    \begin{subfigure}{0.19\textwidth}
         \centering
         \includegraphics[width=\textwidth]{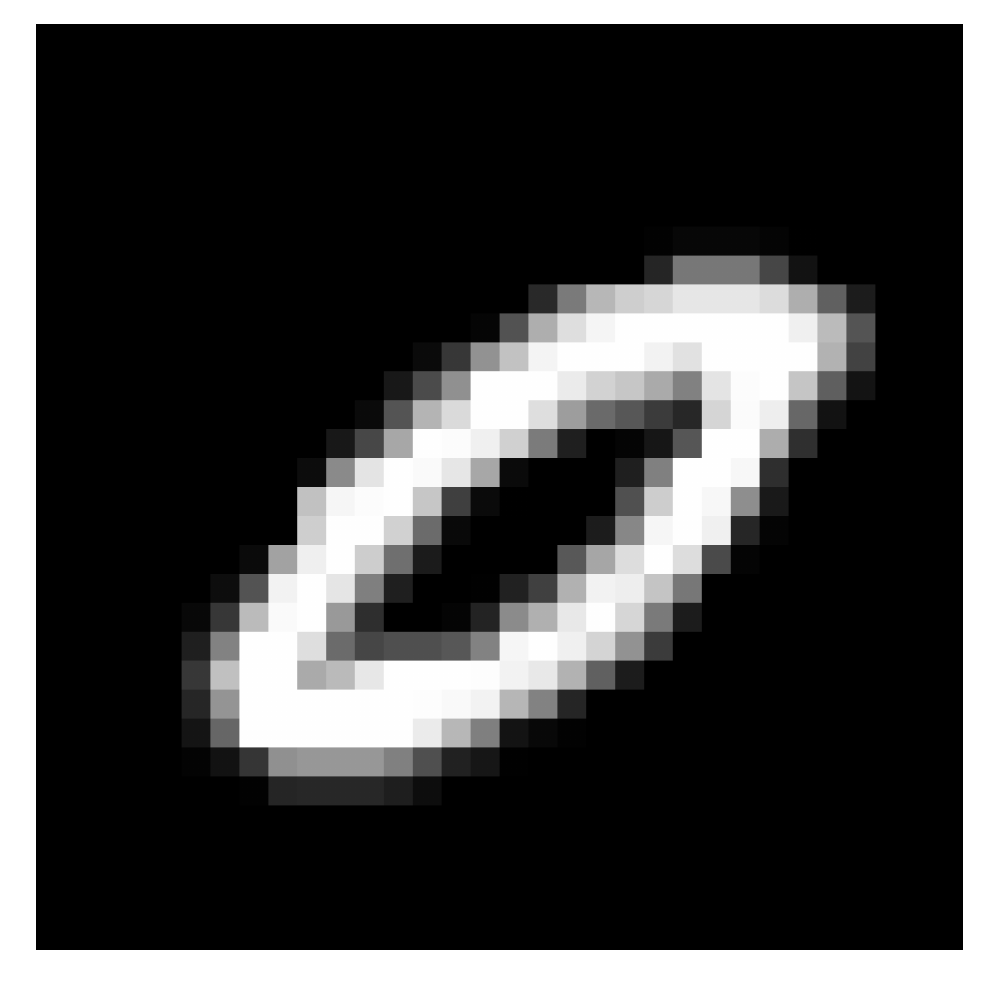}
    \end{subfigure}
    
    \begin{subfigure}{0.19\textwidth}
         \centering
         \includegraphics[width=\textwidth]{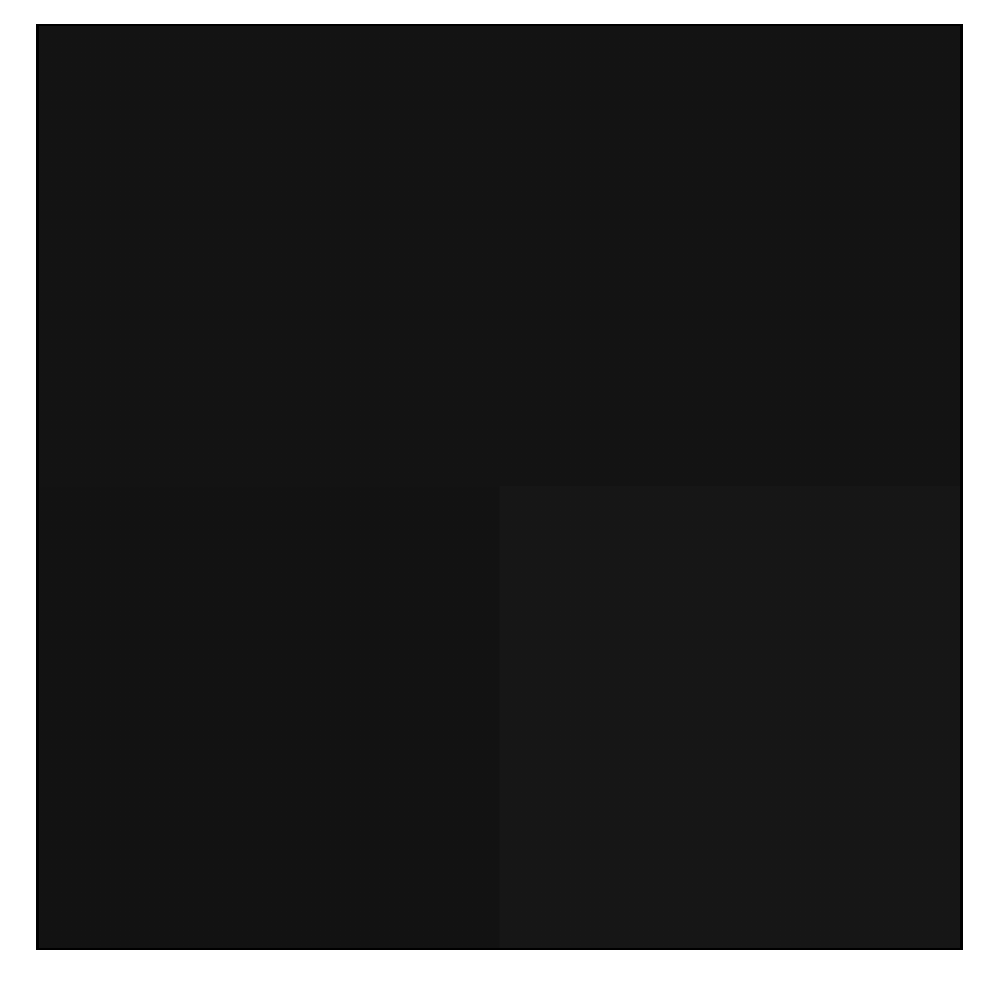}
    \end{subfigure}
    \begin{subfigure}{0.19\textwidth}
         \centering
         \includegraphics[width=\textwidth]{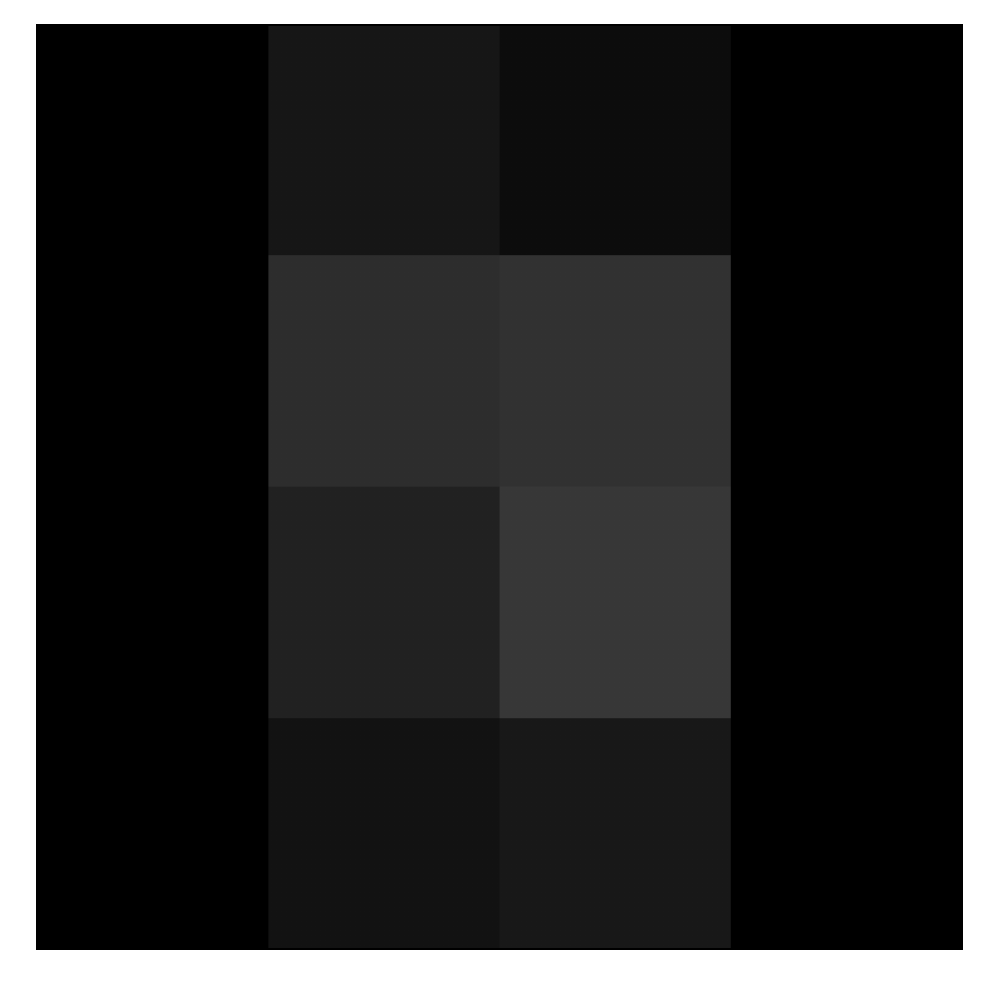}
    \end{subfigure}
        \begin{subfigure}{0.19\textwidth}
         \centering
         \includegraphics[width=\textwidth]{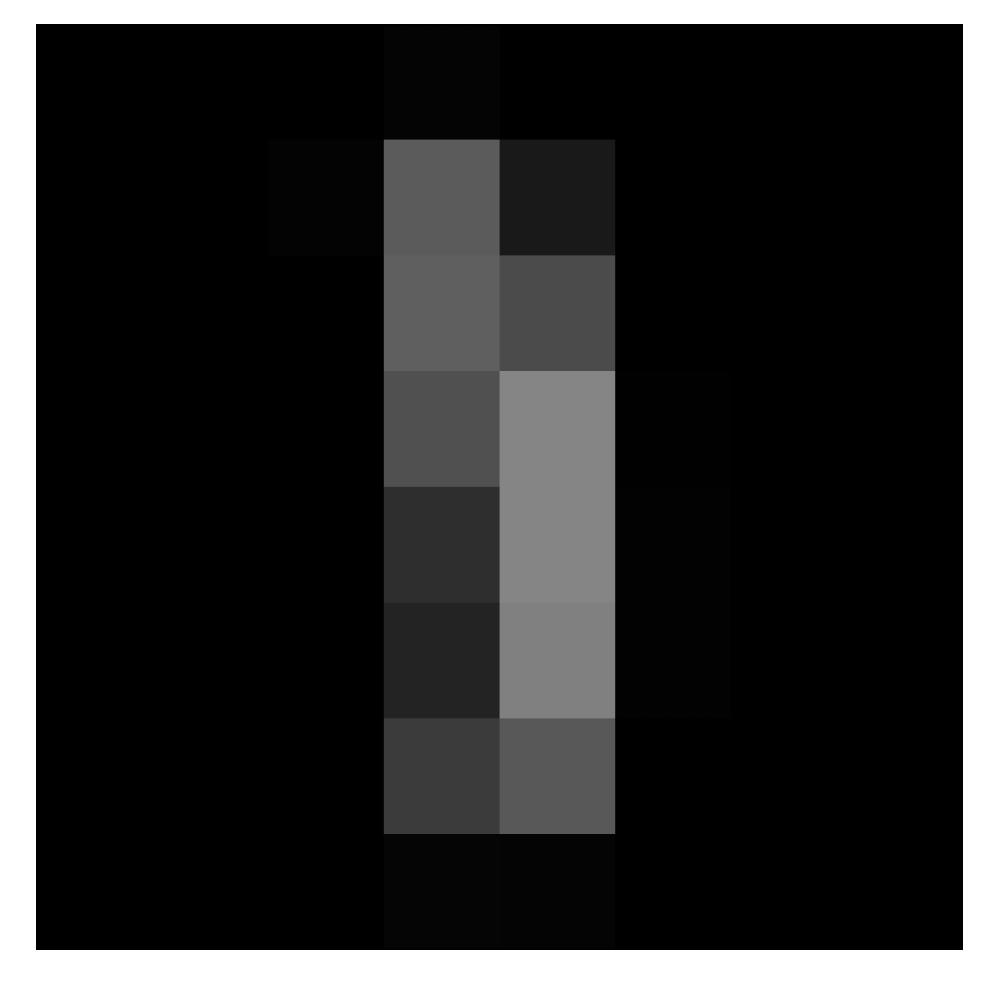}
    \end{subfigure}
    \begin{subfigure}{0.19\textwidth}
         \centering
         \includegraphics[width=\textwidth]{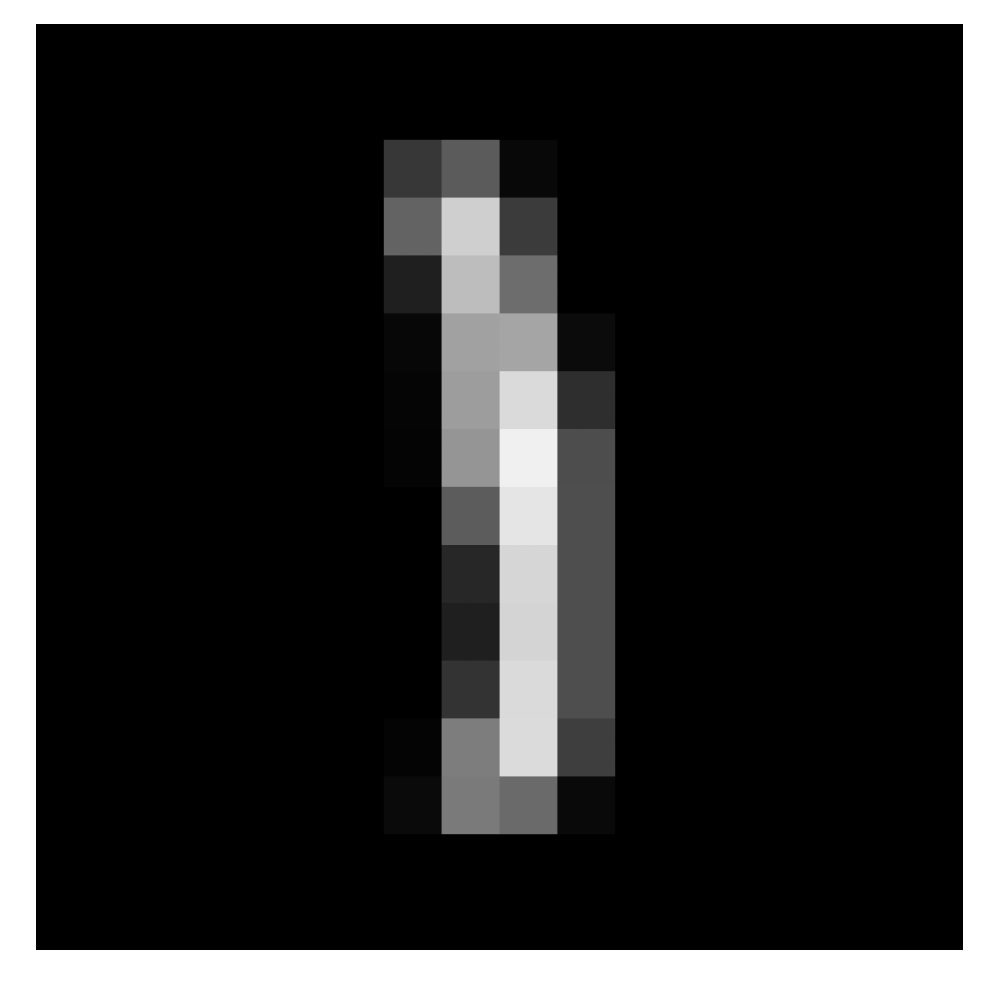}
    \end{subfigure}
        \begin{subfigure}{0.19\textwidth}
         \centering
         \includegraphics[width=\textwidth]{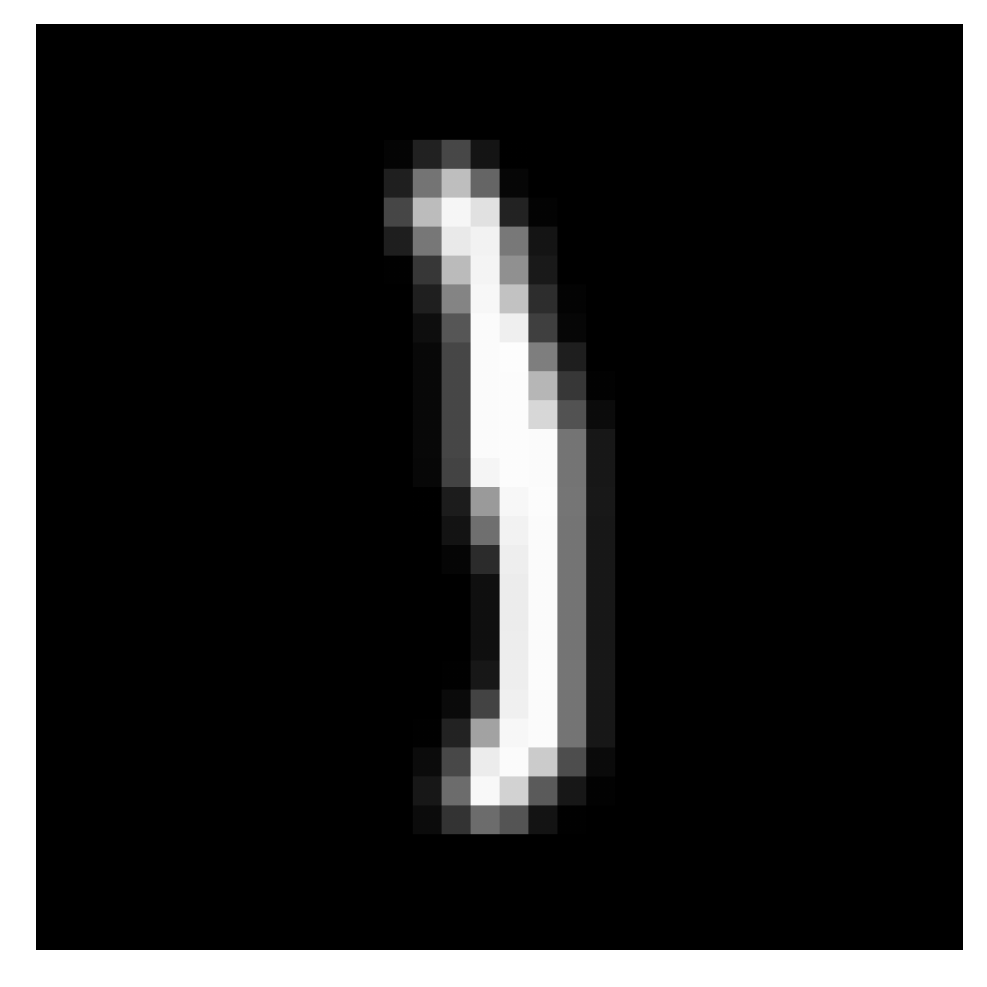}
    \end{subfigure}
    \caption{MNIST data. Row 1 and 2 show digits 0 and 1 respectively for $n\in[1,5]$ from left to right.}
    \label{fig:mnist}
\end{figure*}
\begin{figure*}
    \centering
    \begin{subfigure}{0.19\textwidth}
         \centering
         \includegraphics[width=\textwidth]{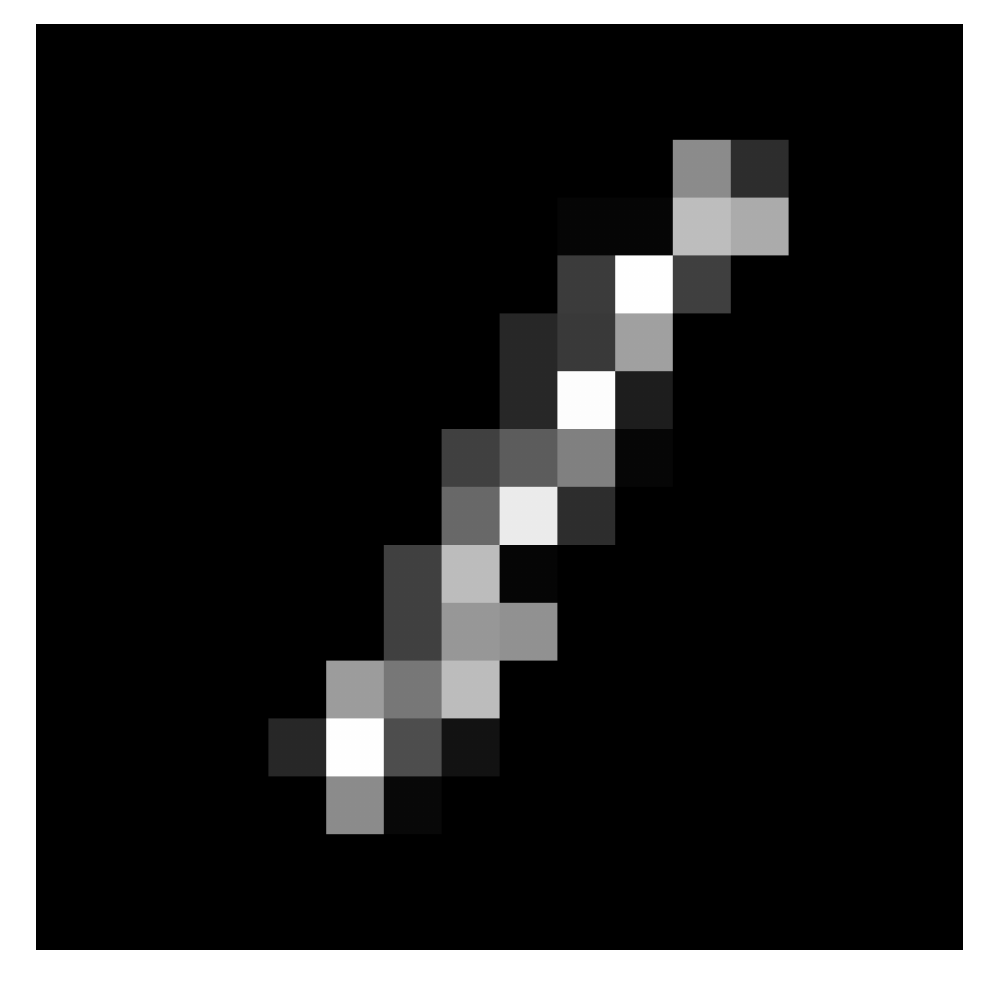}
         \caption{Shot Noise}
    \end{subfigure}
    \begin{subfigure}{0.19\textwidth}
        \centering
        \includegraphics[width=\textwidth]{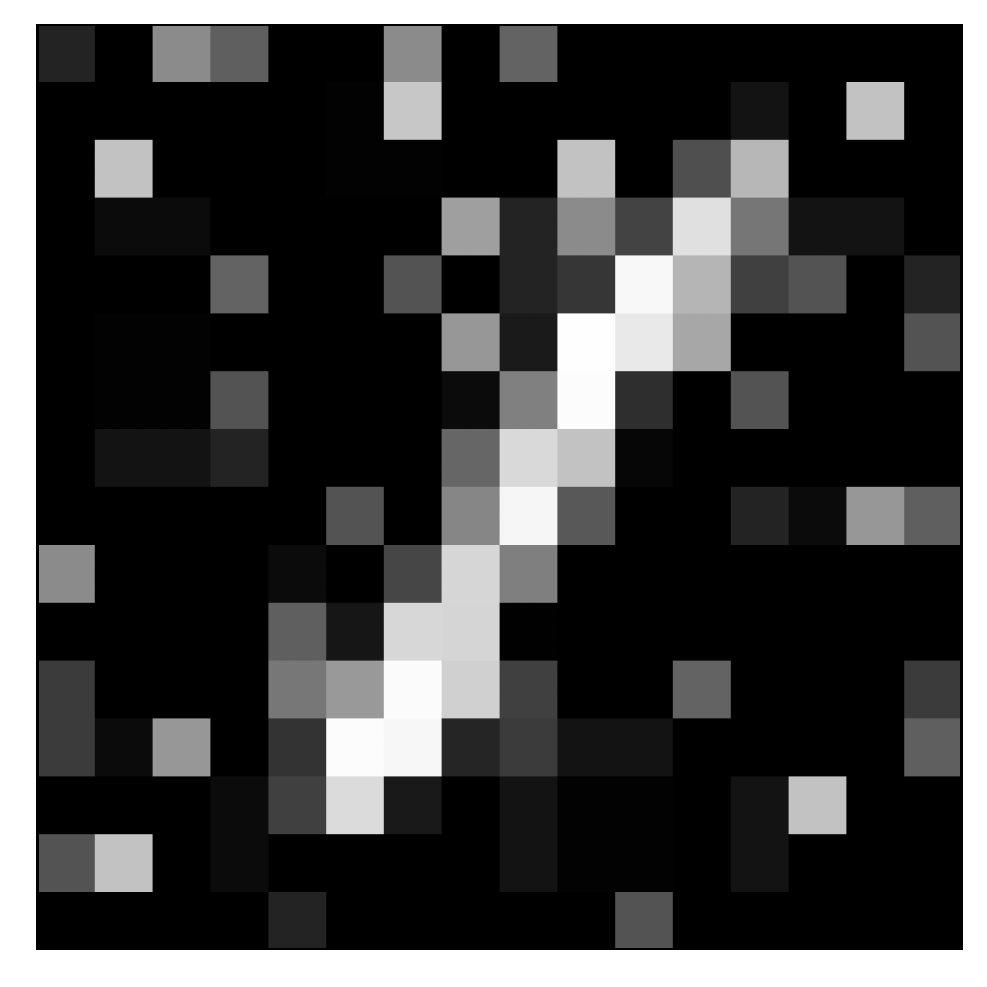}
        \caption{Impulse Noise}
    \end{subfigure}
    \begin{subfigure}{0.19\textwidth}
         \centering
         \includegraphics[width=\textwidth]{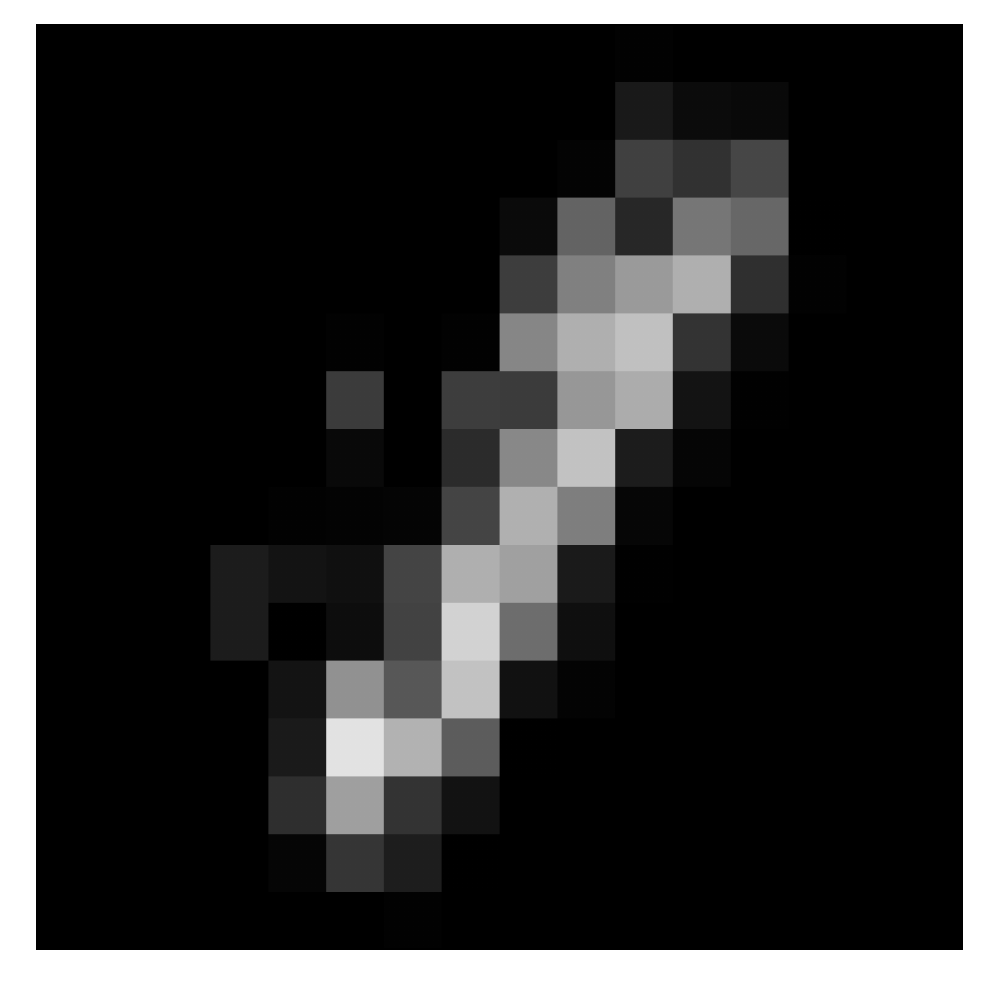}
        \caption{Glass Blur}
    \end{subfigure}
    \begin{subfigure}{0.19\textwidth}
         \centering
         \includegraphics[width=\textwidth]{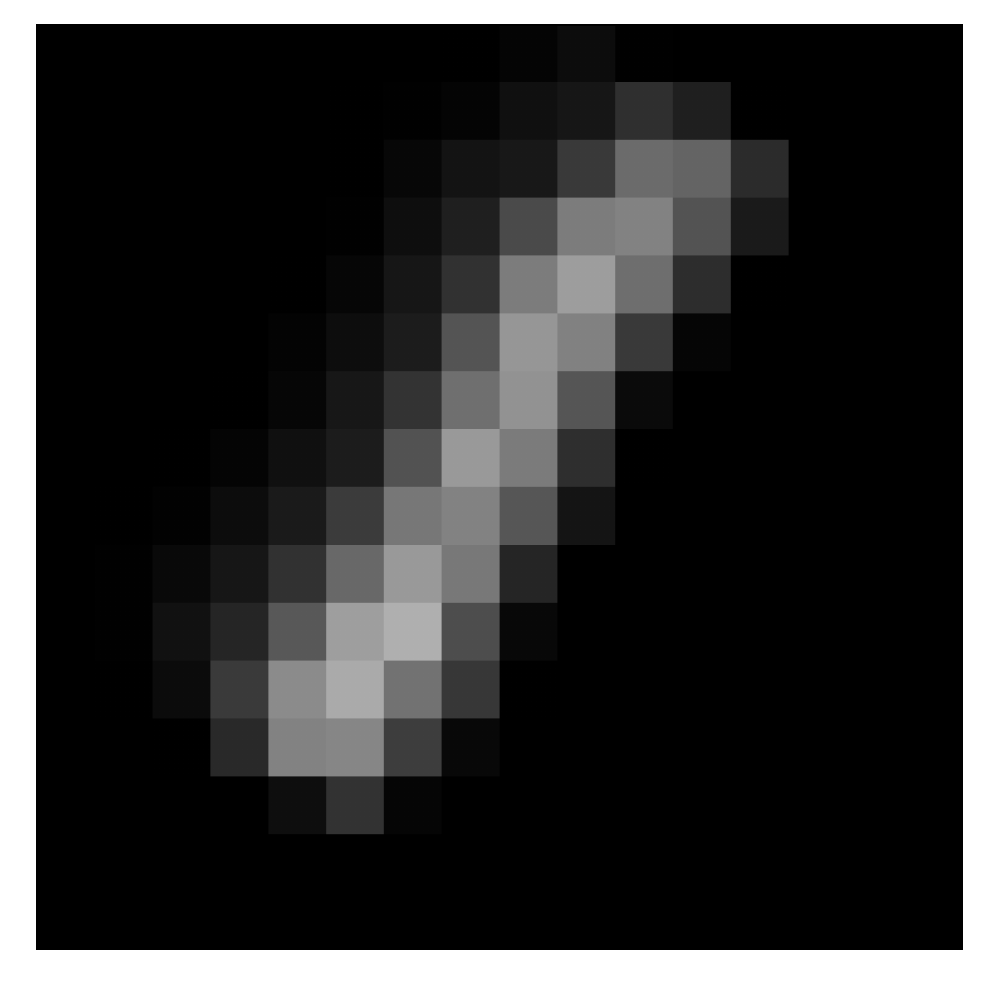}
         \caption{Motion Blur}
    \end{subfigure}
    \begin{subfigure}{0.19\textwidth}
         \centering
         \includegraphics[width=\textwidth]{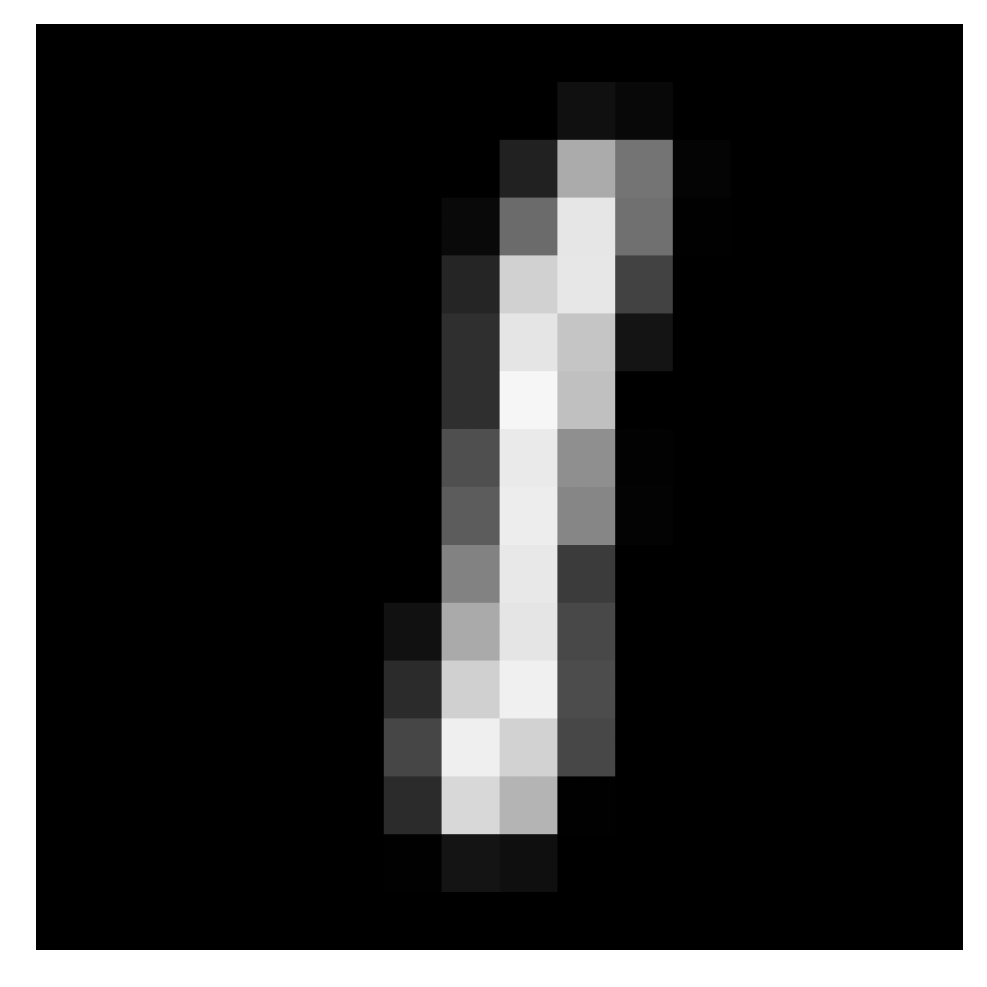}
         \caption{Shear}
    \end{subfigure}
    
    \begin{subfigure}{0.19\textwidth}
         \centering
         \includegraphics[width=\textwidth]{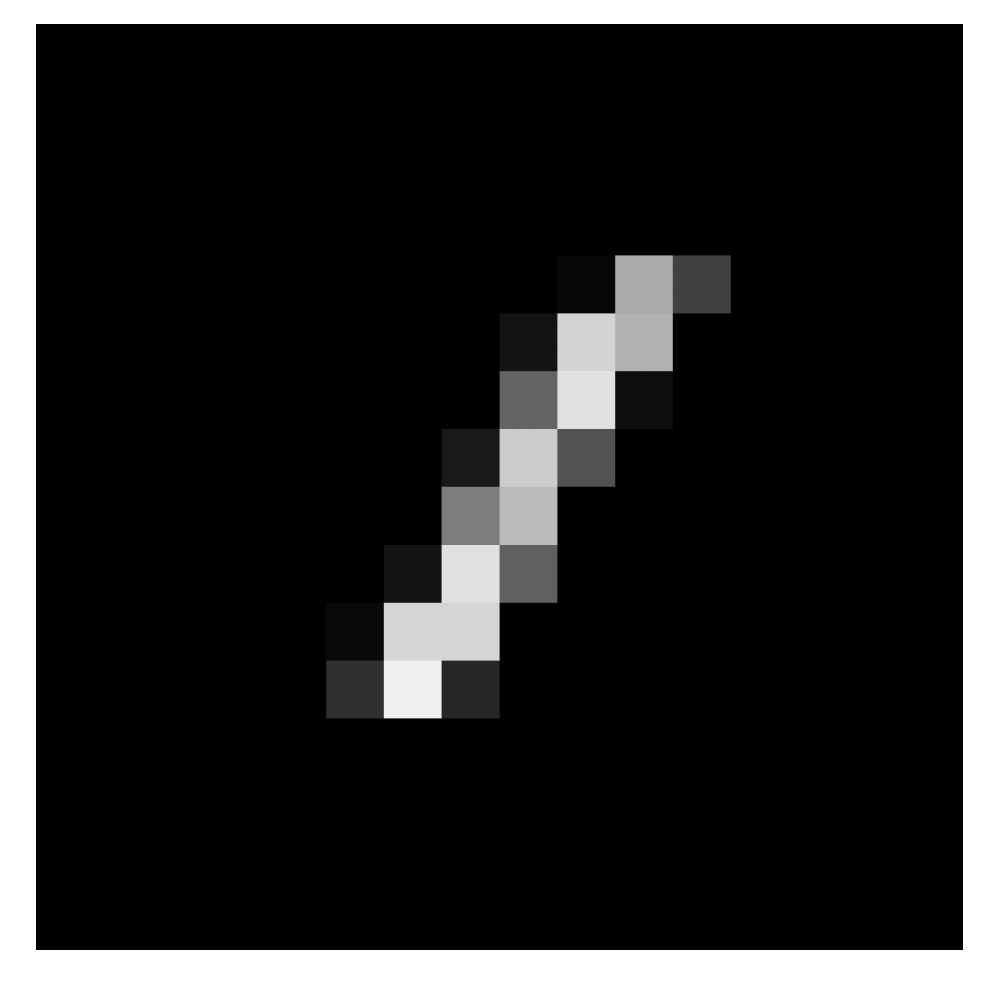}
         \caption{Scale}
    \end{subfigure}
    \begin{subfigure}{0.19\textwidth}
         \centering
         \includegraphics[width=\textwidth]{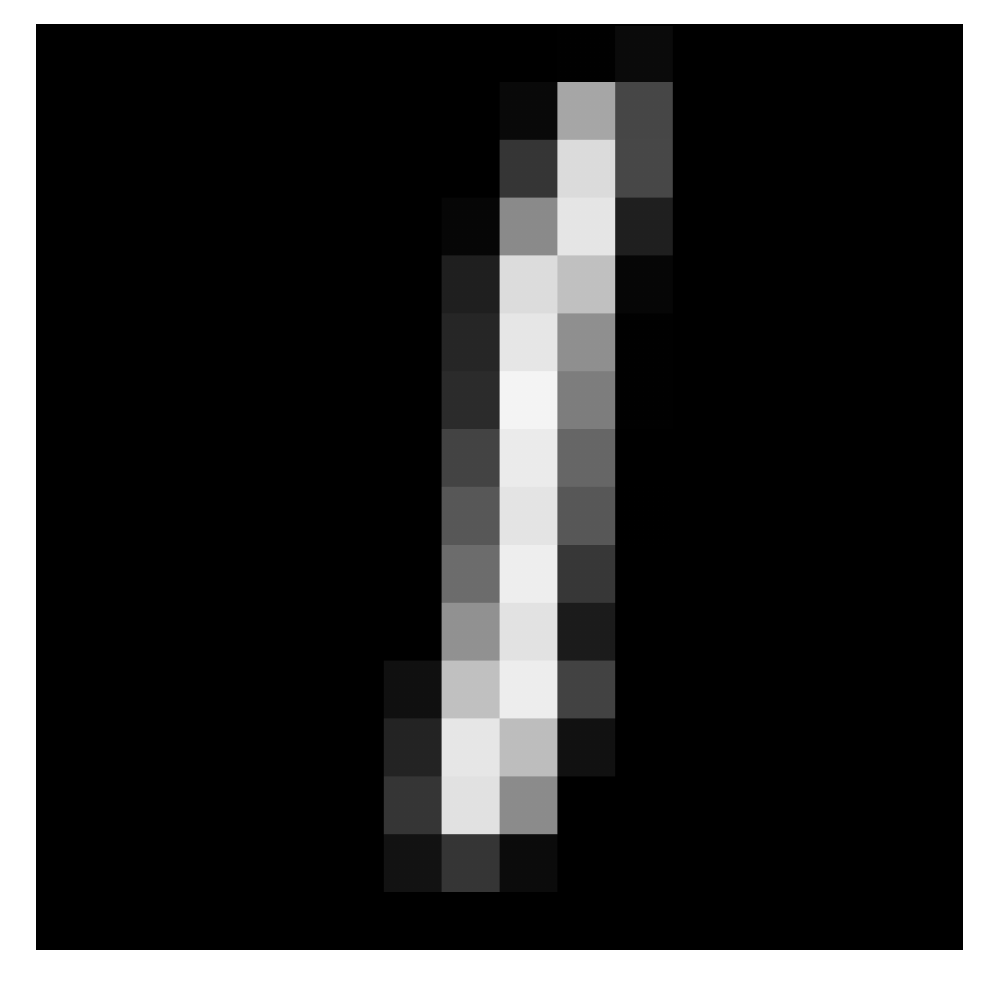}
         \caption{Rotate}
    \end{subfigure}
        \begin{subfigure}{0.19\textwidth}
         \centering
         \includegraphics[width=\textwidth]{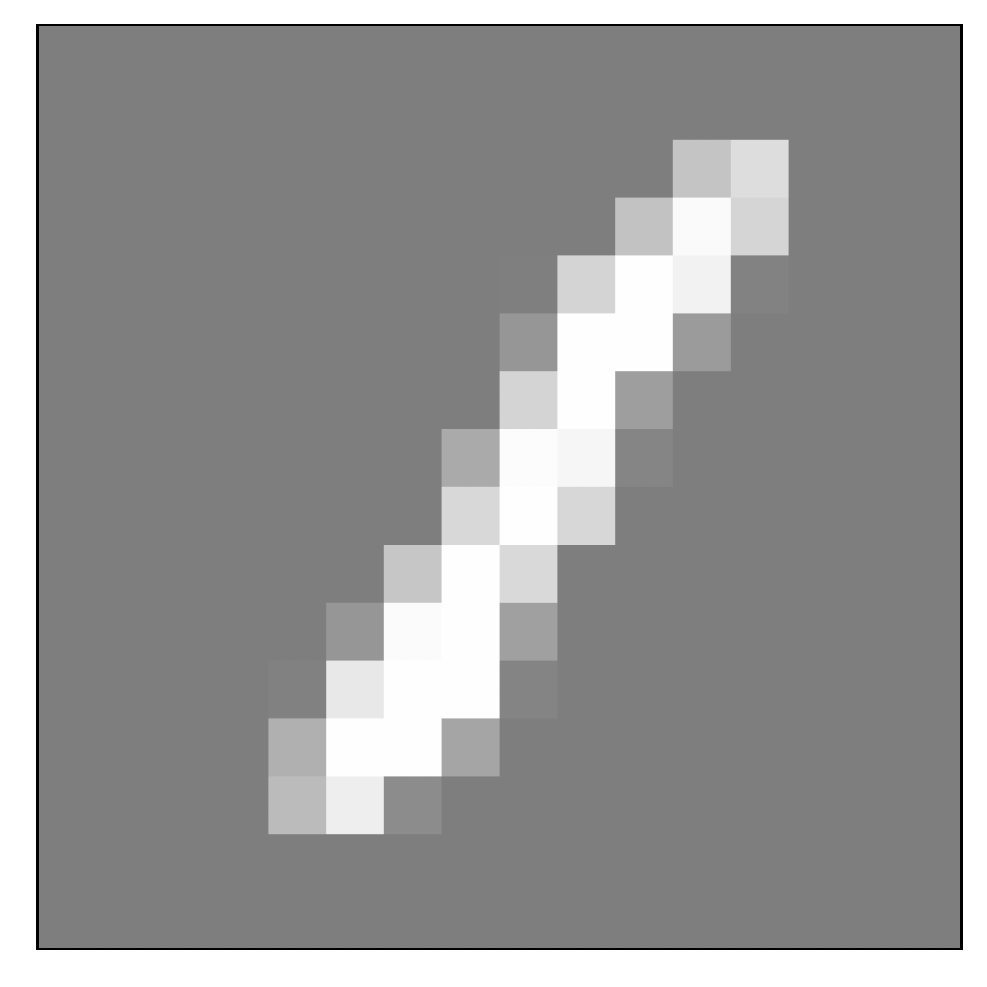}
         \caption{Brightness}
    \end{subfigure}
    \begin{subfigure}{0.19\textwidth}
         \centering
         \includegraphics[width=\textwidth]{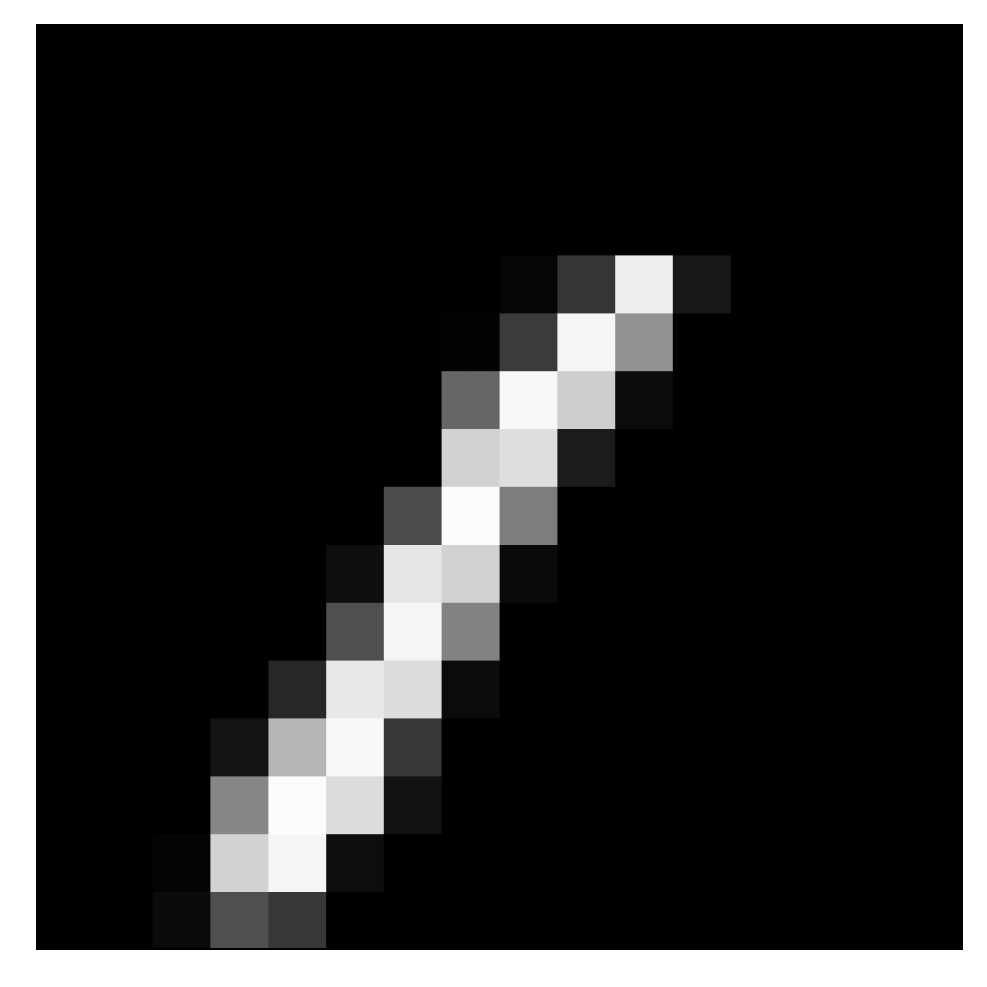}
         \caption{Translate}
    \end{subfigure}
        \begin{subfigure}{0.19\textwidth}
         \centering
         \includegraphics[width=\textwidth]{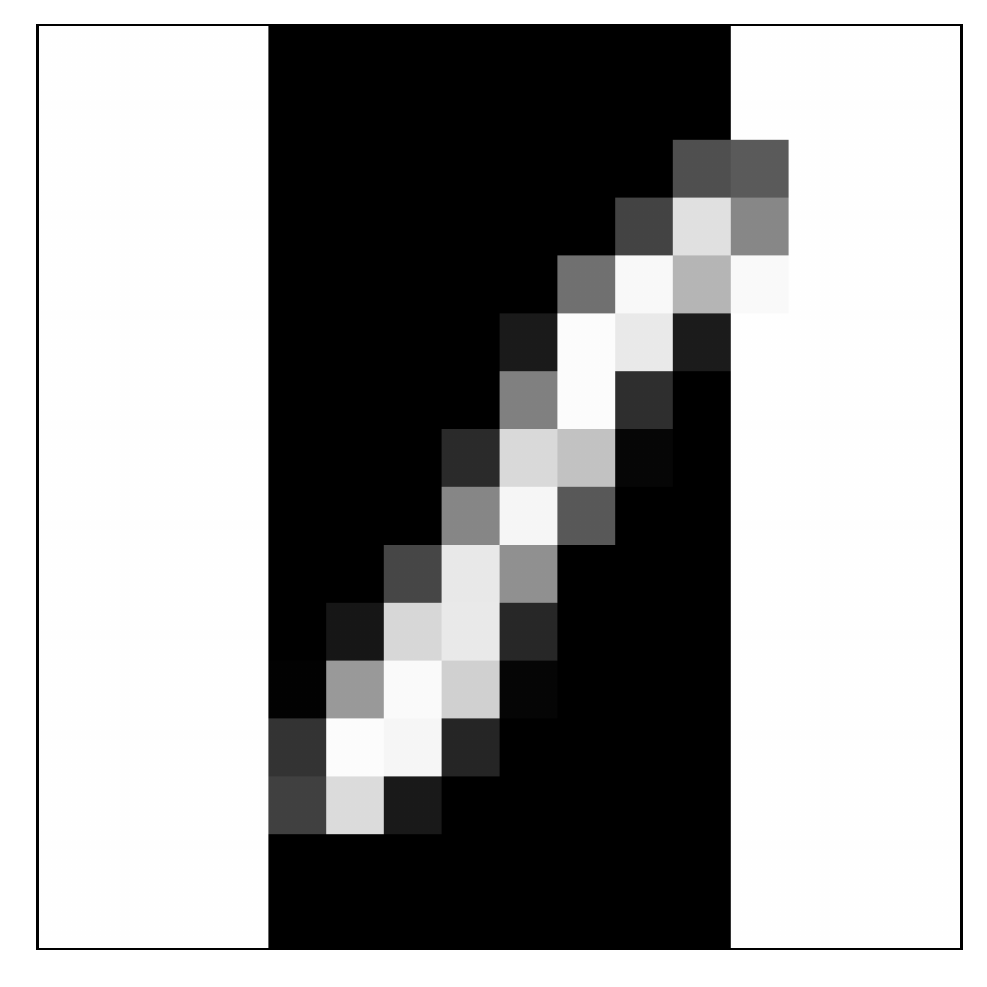}
         \caption{Stripe}
    \end{subfigure}
    
    \begin{subfigure}{0.19\textwidth}
         \centering
         \includegraphics[width=\textwidth]{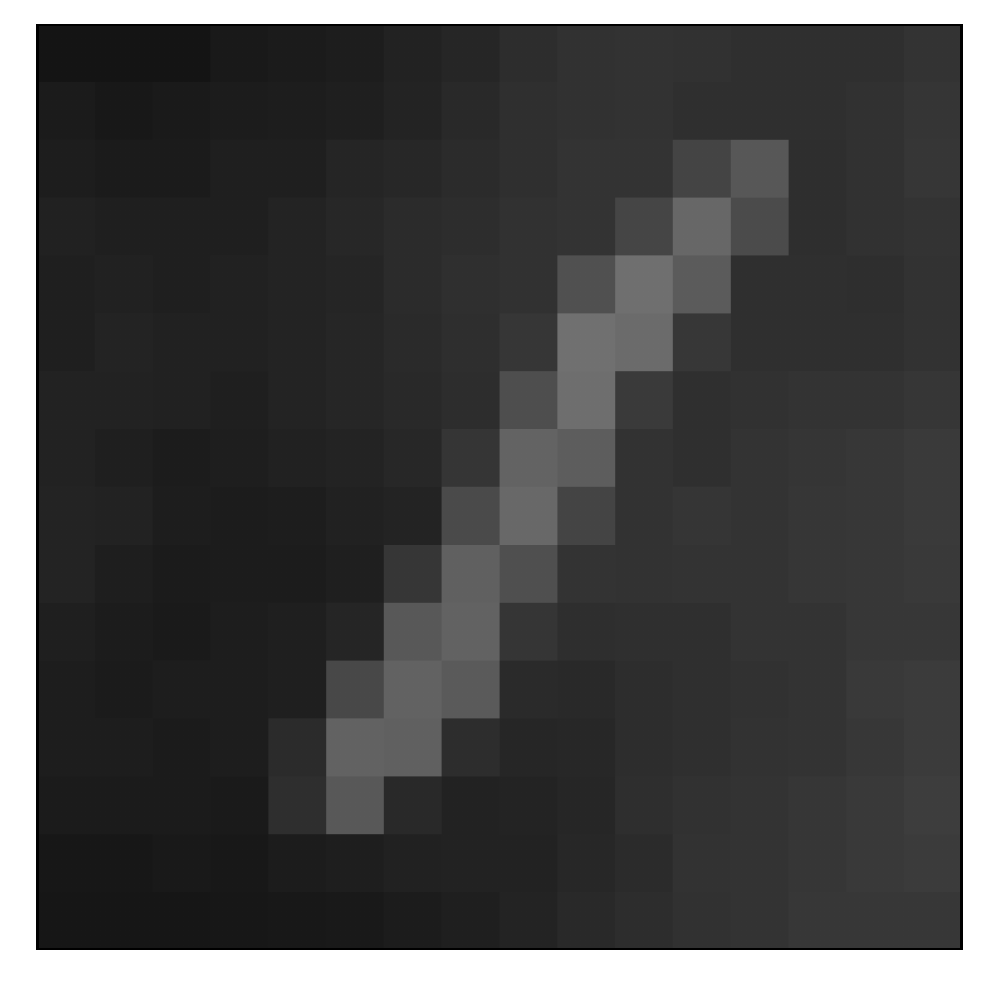}
         \caption{Fog}
    \end{subfigure}
    \begin{subfigure}{0.19\textwidth}
         \centering
         \includegraphics[width=\textwidth]{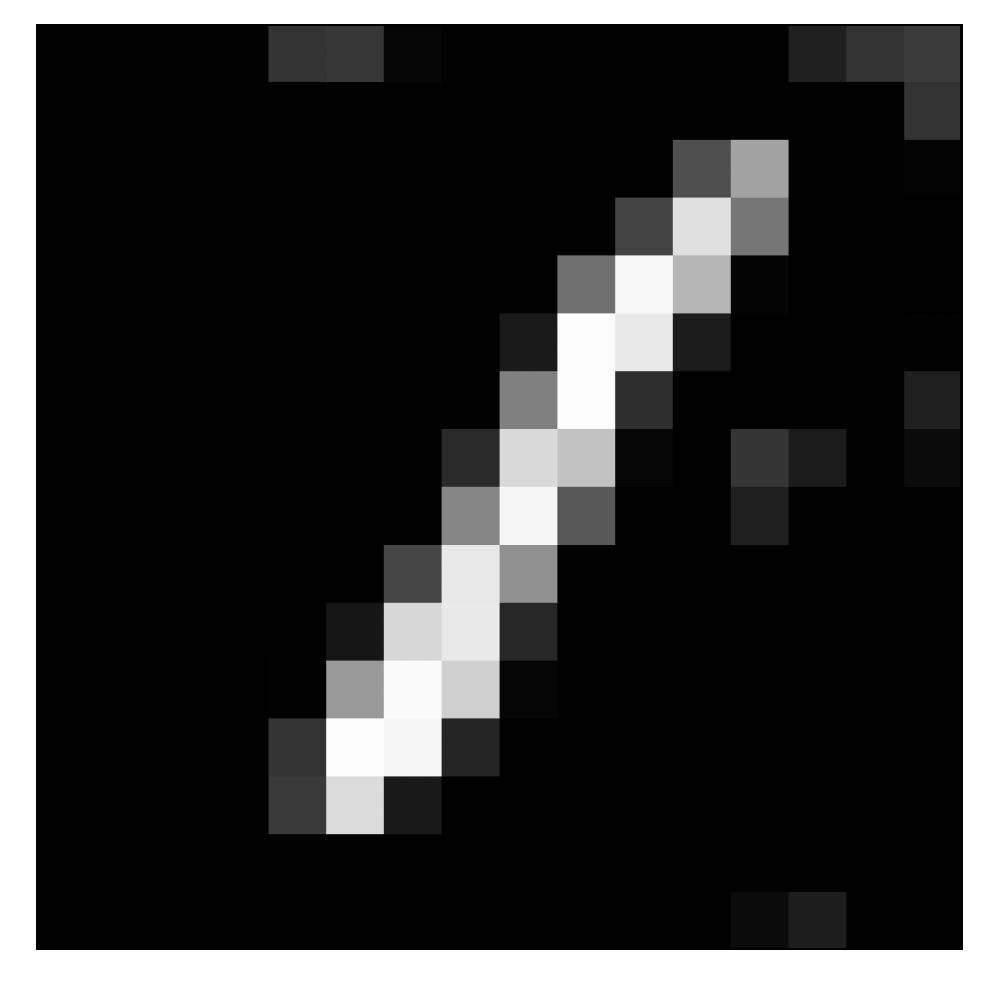}
         \caption{Spatter}
    \end{subfigure}
        \begin{subfigure}{0.19\textwidth}
         \centering
         \includegraphics[width=\textwidth]{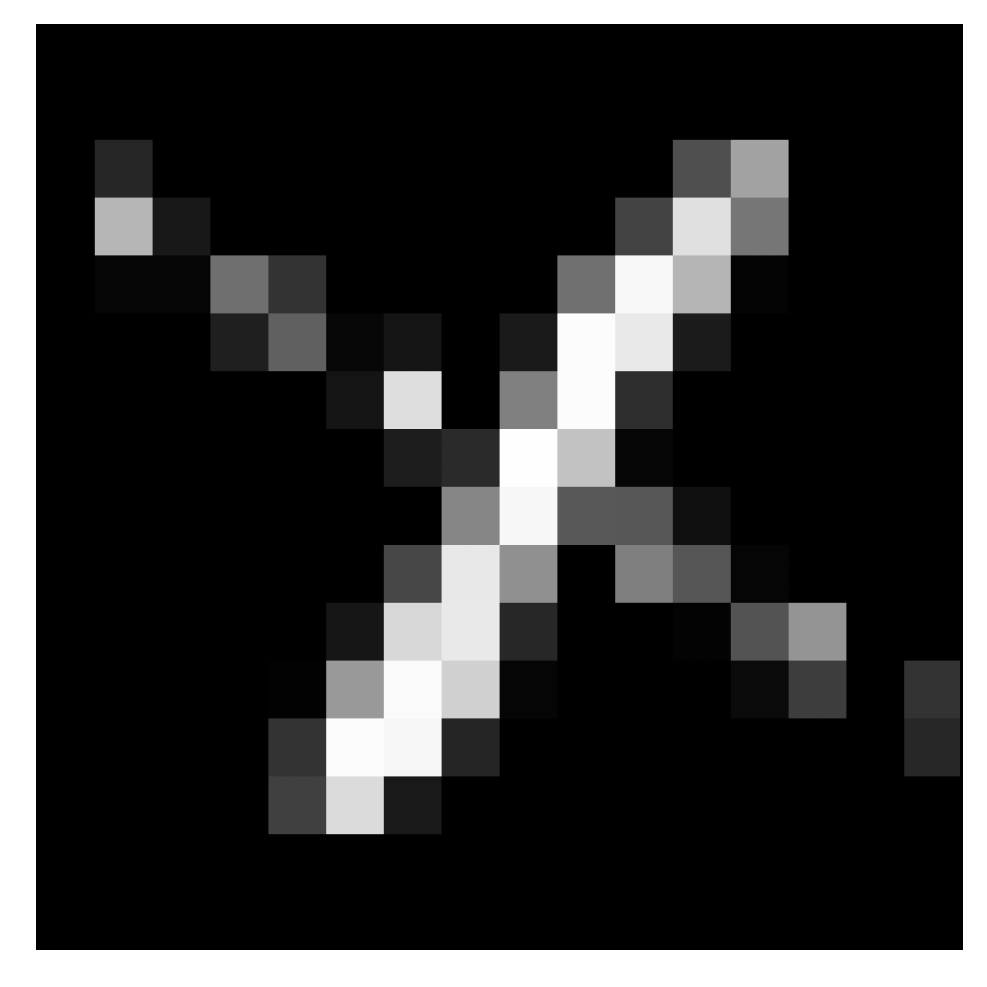}
         \caption{Dotted Line}
    \end{subfigure}
    \begin{subfigure}{0.19\textwidth}
         \centering
         \includegraphics[width=\textwidth]{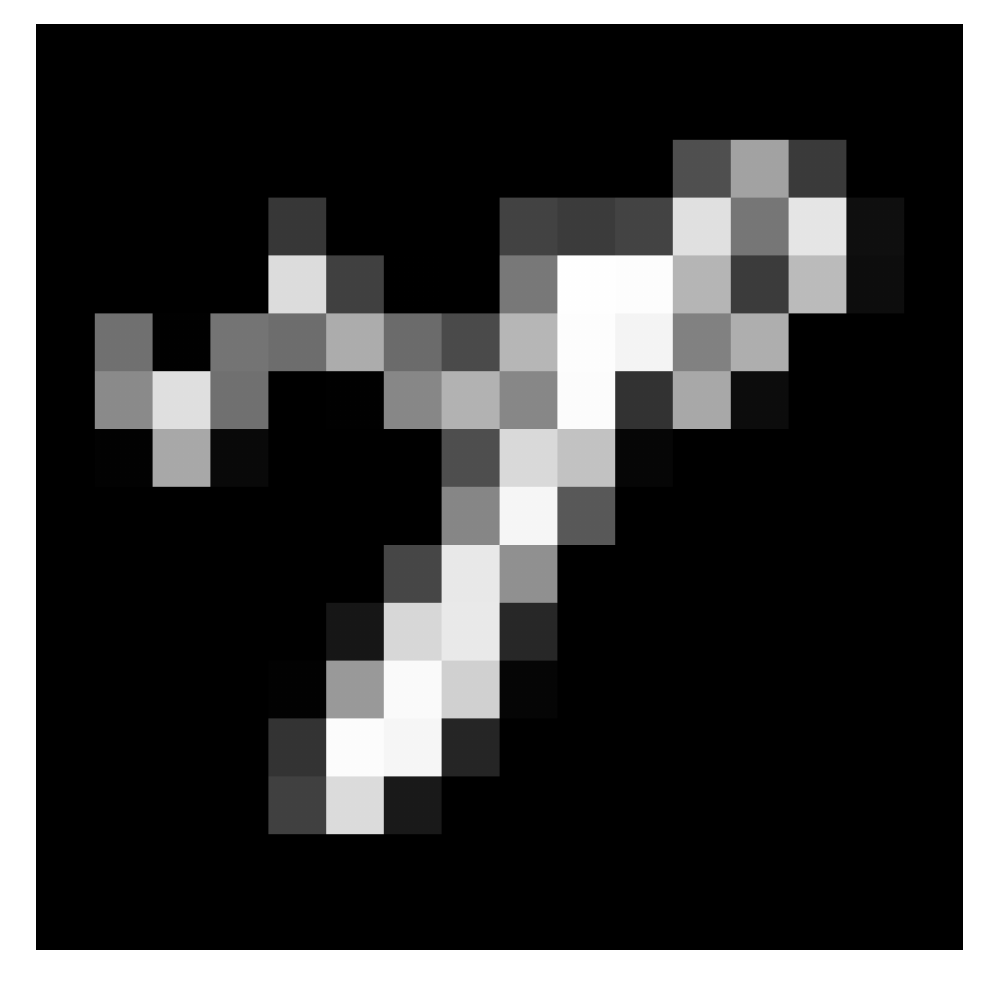}
         \caption{ZigZag}
    \end{subfigure}
        \begin{subfigure}{0.19\textwidth}
         \centering
         \includegraphics[width=\textwidth]{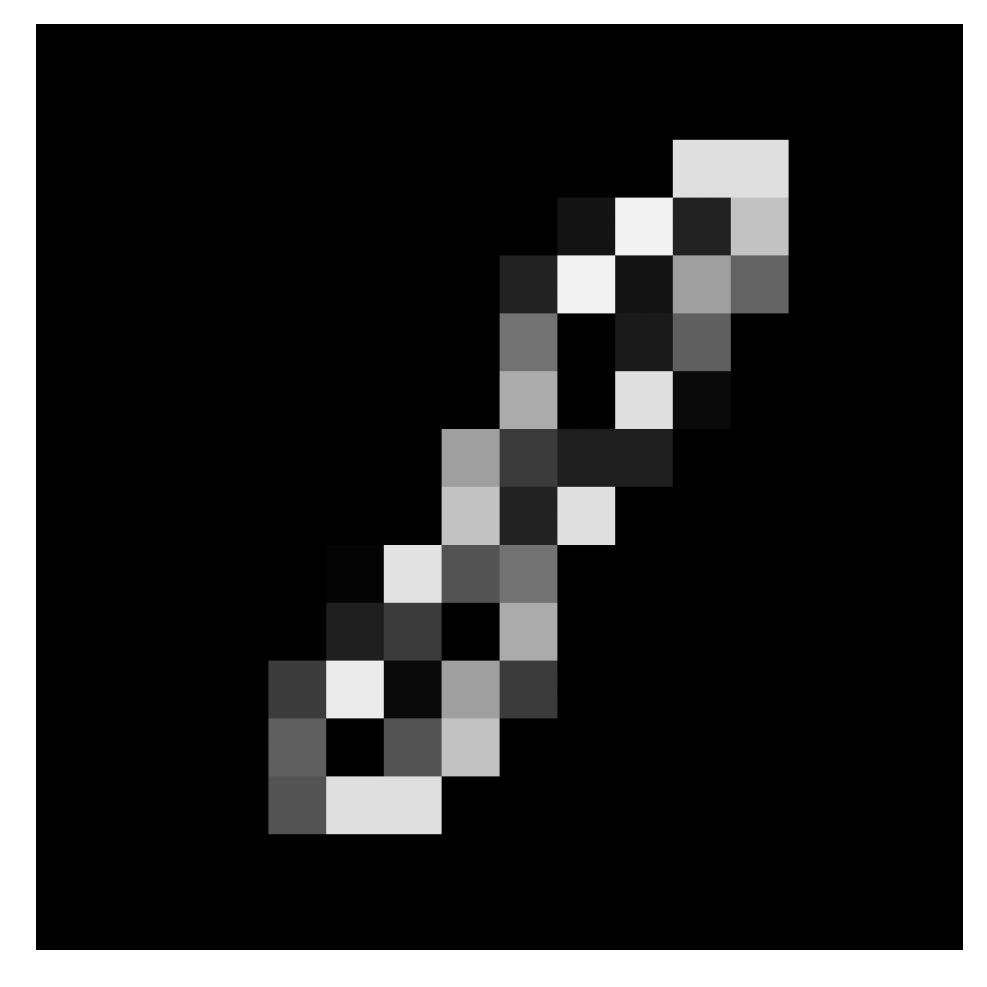}
         \caption{Canny Edges}
    \end{subfigure}
    \caption{Corrupted MNIST data. Example of all 15 corruptions. $n=4$.}
    \label{fig:mnist_noise}
\end{figure*}

This is the famous dataset of handwritten digits \cite{lecun2010mnist}. We have used this for both binary (0 and 1) and multi-class (0, 1 and 2) classification. The original images are of $28\times28$ dimension, which first needs to be squared into $2^n\times2^n$. Bilinear interpolation is used to transform the data for different $n$. There also exists another version of this data which contains 15 different corruption variations \cite{mu2019mnist}. We also classify these corrupted datasets. Fig \ref{fig:mnist} shows the  MNIST images, and Fig \ref{fig:mnist_noise} shows the corrupted images.

\subsection{\texorpdfstring{$2\times2$}{2 x 2} Color Images}
\begin{figure*}
    \centering
    \begin{subfigure}{0.15\textwidth}
         \centering
         \includegraphics[width=\textwidth]{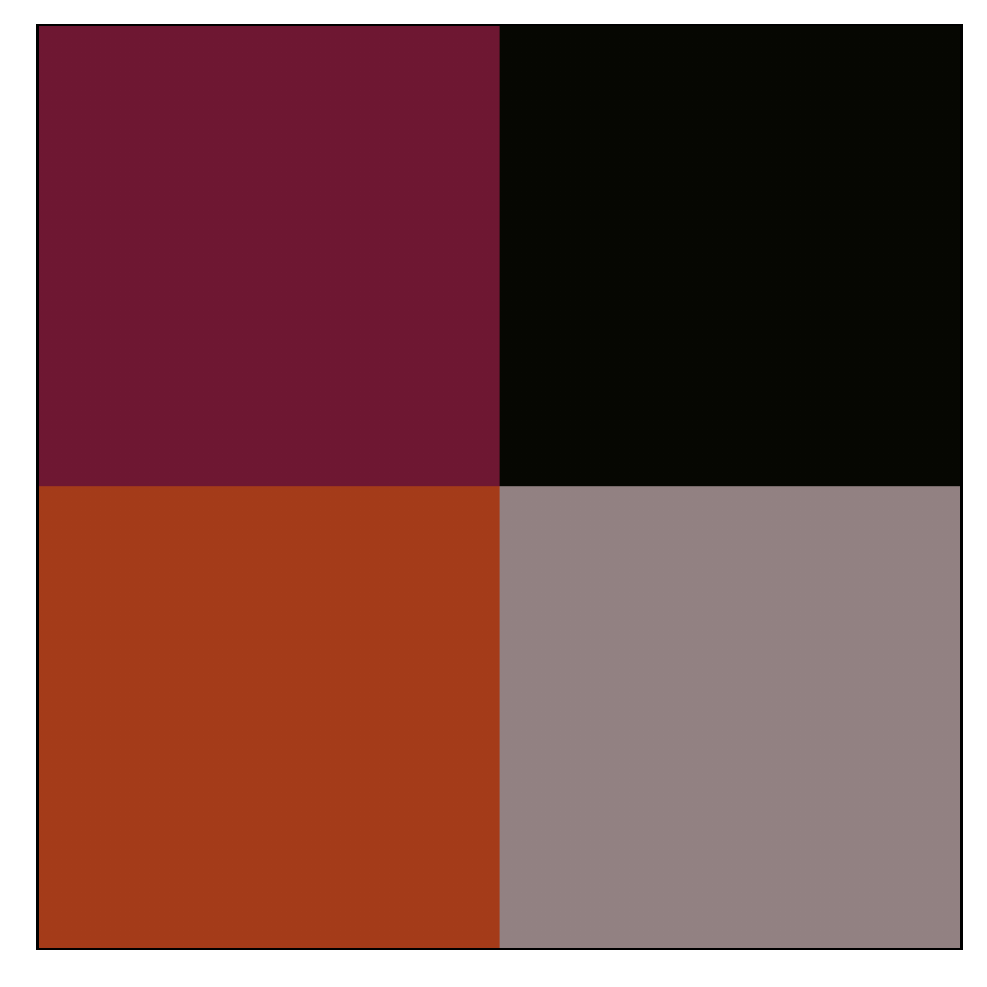}
         \caption{0 Class}
    \end{subfigure}
        \begin{subfigure}{0.15\textwidth}
         \centering
         \includegraphics[width=\textwidth]{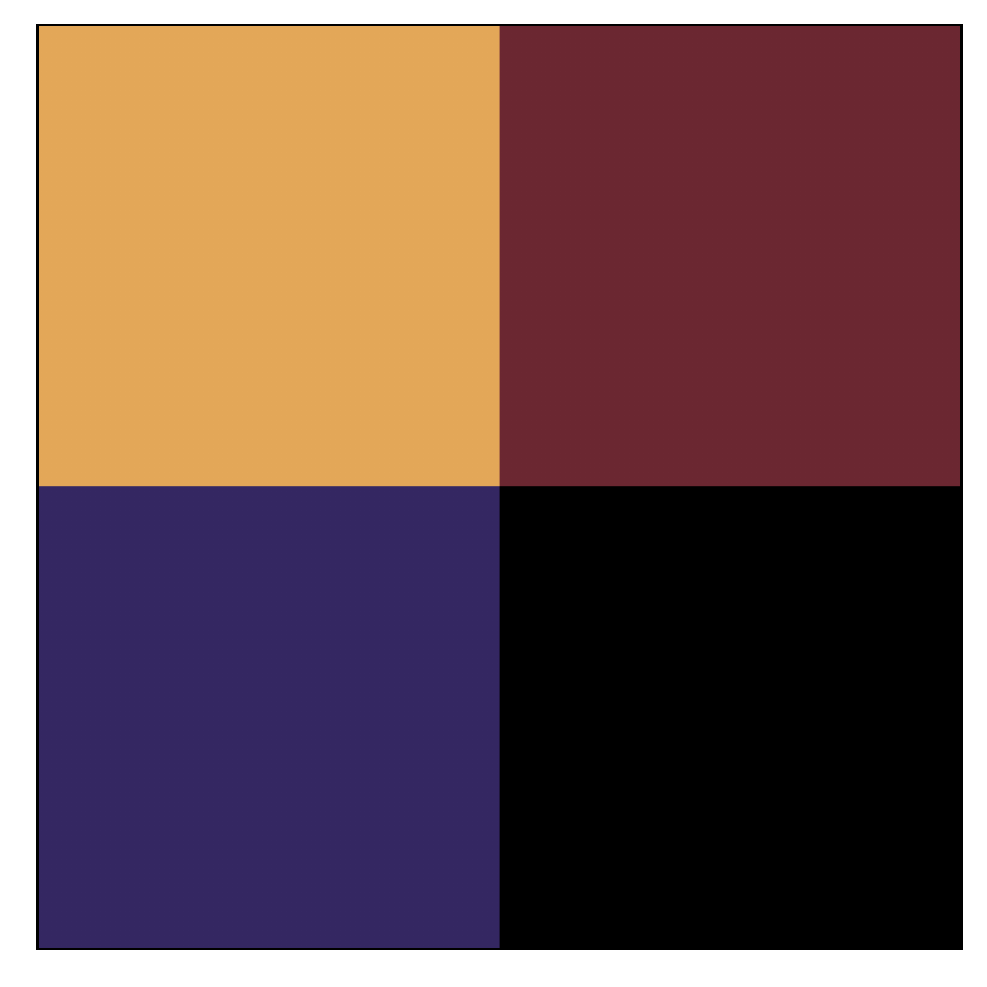}
         \caption{1 Class, 0 shade}
    \end{subfigure}
    \begin{subfigure}{0.15\textwidth}
        \centering
        \includegraphics[width=\textwidth]{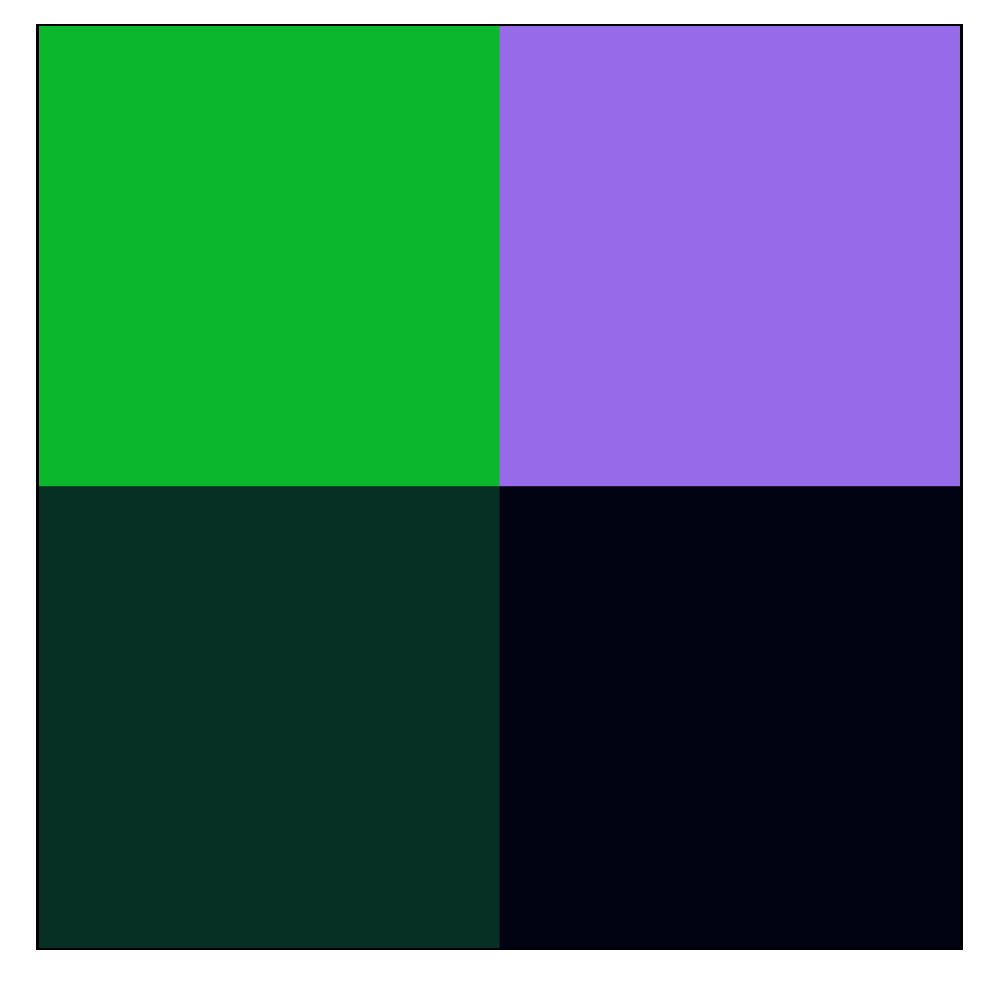}
        \caption{1 Class, 20 shade}
    \end{subfigure}
    \begin{subfigure}{0.15\textwidth}
         \centering
         \includegraphics[width=\textwidth]{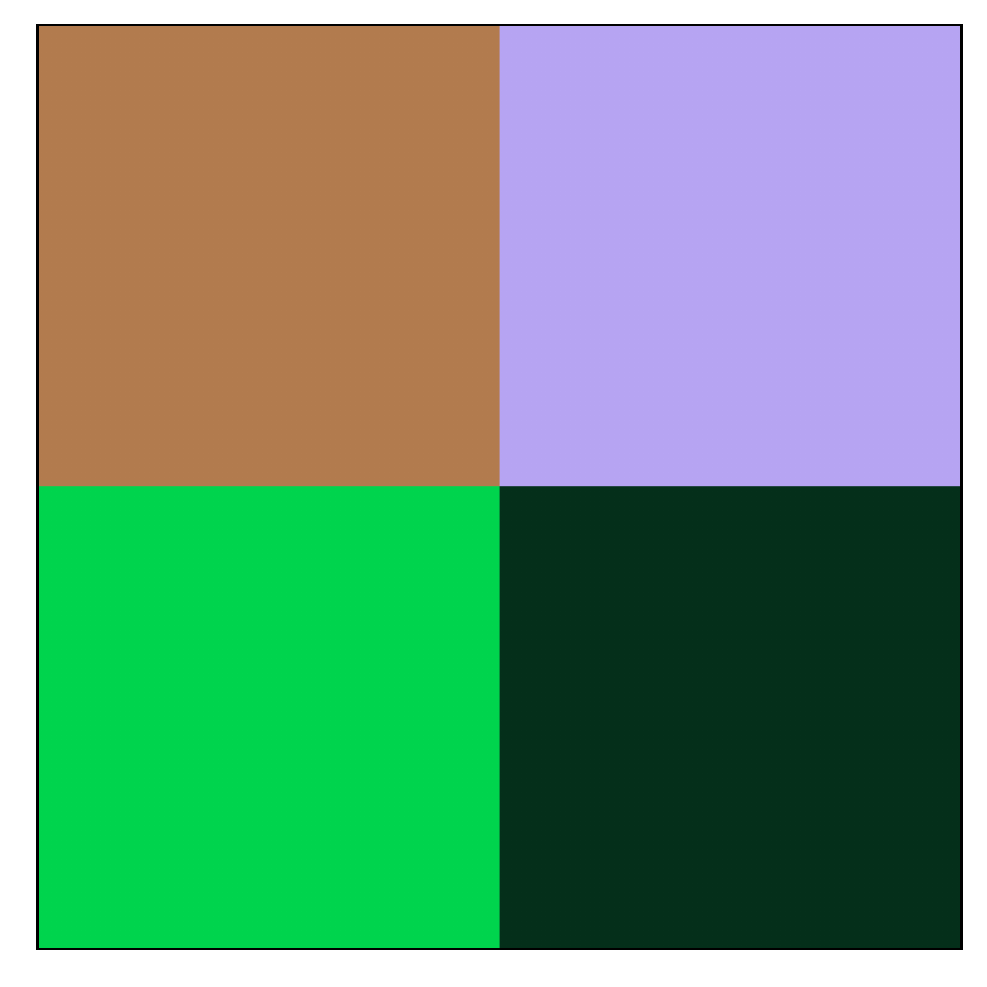}
        \caption{1 Class, 50 shade}
    \end{subfigure}
    \begin{subfigure}{0.15\textwidth}
         \centering
         \includegraphics[width=\textwidth]{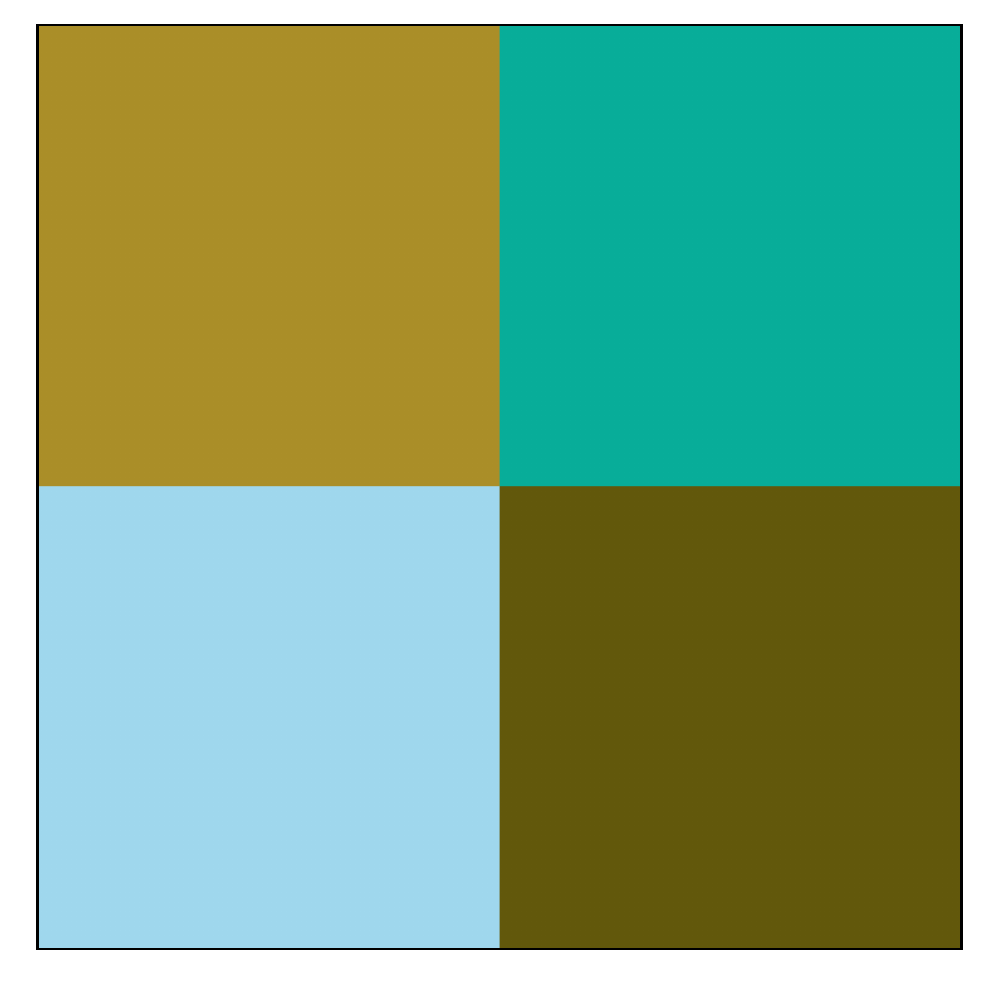}
         \caption{1 Class, 100 shade}
    \end{subfigure}
    \begin{subfigure}{0.15\textwidth}
         \centering
         \includegraphics[width=\textwidth]{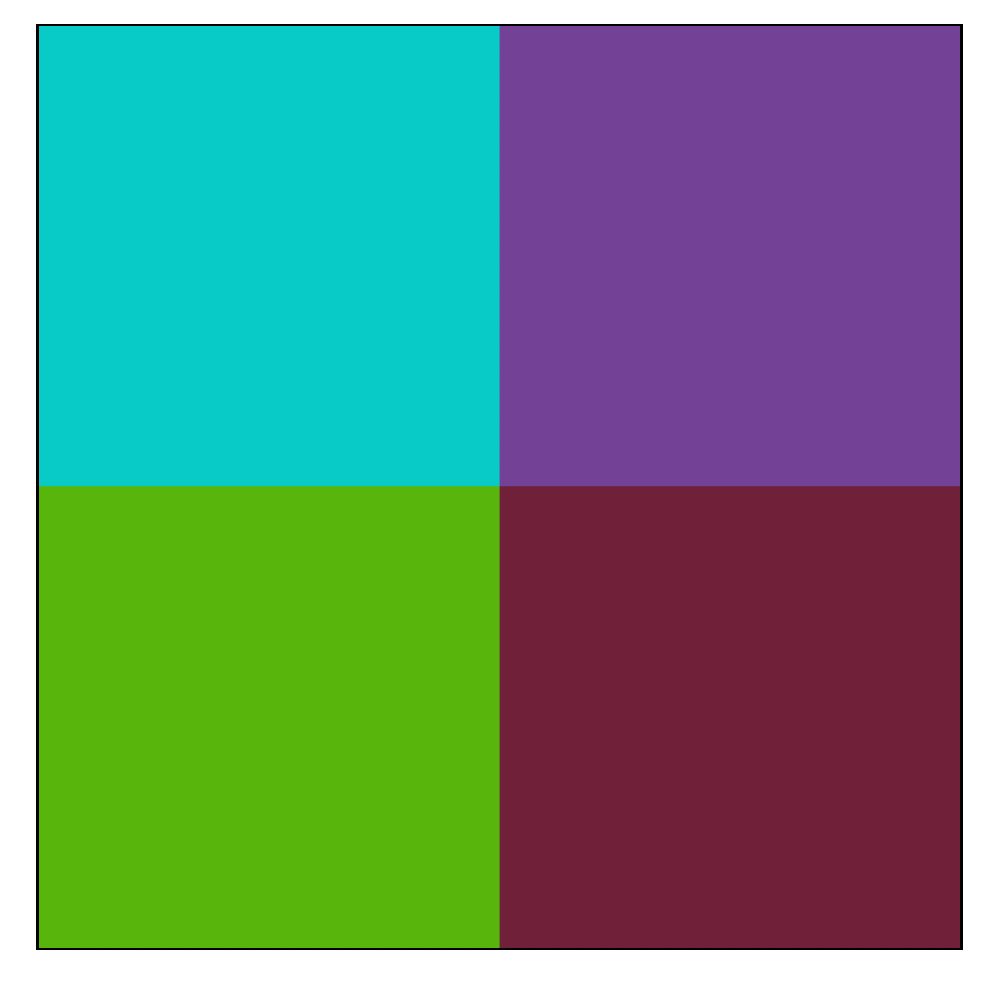}
         \caption{1 Class, 200 shade}
    \end{subfigure}
    \caption{$2\times2$ Color images with different shades of the positive class.}
    \label{fig:22_color}
\end{figure*}

This is randomly generated $2\times2$ colour image data for classification. The image has 4 pixels of random colours. For positive class we change the pixel values of the $4^{th}$ pixel to $(0,0,0)$. This makes the pixel black. The classification problem is then to differentiate between images with and without a black pixel. The pixel can also be modified to have dark shades instead of absolute 0 values. Fig \ref{fig:22_color} shows example images for different shades.

\section{Implementation Details}
\label{implementation}
We use PennyLane \cite{https://doi.org/10.48550/arxiv.1811.04968} to simulate the circuits. JAX \cite{jax2018github} is combined with PennyLane as a high-performance simulator and to utilise the GPU. The optimisation library optax \cite{optax2020github} is used for optimising the classifier. We use the Adam optimiser \cite{https://doi.org/10.48550/arxiv.1412.6980} with a 0.1 step size for optimisation for 250 epochs unless otherwise mentioned. We use 5 layers of the VQC and 1 layer of the AC. We also use scikit-learn \cite{scikit-learn} for classical processing. All simulations were done with `None' shots of the PennyLane device. This gives analytic results. We use different sizes of training datasets and a fixed 1000 data points for the validation data.

\section{Results}
\label{results}
\subsection{BAS}
Table \ref{tab:res_bas} shows accuracies on the validation set using different training dataset sizes and $n$ values for the BAS data.
\begin{table}[!htbp]
\centering
\caption{\textbf{Validation Set Accuracy using the VQC and the AC on FRQI representation for BAS data}}
\label{tab:res_bas}
\begin{tabular}{@{}cccccc@{}}
\toprule
\textbf{\begin{tabular}[c]{@{}c@{}}Training Set\\ Size\end{tabular}} & \textbf{Classifier} & \multicolumn{4}{c}{\textbf{$n$}} \\ \midrule
\textbf{} & \textbf{} & \textit{\textbf{1}} & \textit{\textbf{2}} & \textit{\textbf{3}} & \textit{\textbf{4}} \\ \midrule
\multirow{2}{*}{\textbf{100}} & \textit{\textbf{VQC}} & 1.0 & 1.0 & 0.955 & 0.993 \\
 & \textit{\textbf{AC}} & 1.0 & 0.818 & 0.820 & 0.889 \\ \midrule
\multirow{2}{*}{\textbf{200}} & \textit{\textbf{VQC}} & 1.0 & 1.0 & 0.997 & 0.990 \\
 & \textit{\textbf{AC}} & 1.0 & 0.929 & 0.881 & 0.732 \\ \midrule
\multirow{2}{*}{\textbf{500}} & \textit{\textbf{VQC}} & 1.0 & 1.0 & 0.993 & 0.991 \\
 & \textit{\textbf{AC}} & 1.0 & 0.858 & 0.842 & 0.866 \\ \bottomrule
\end{tabular}
\end{table}

\subsection{MNIST}

Table \ref{tab:res_mnist} shows accuracies on the validation set using different training dataset sizes and $n$ values for the MNIST data. For AC we have used 100 epochs for $n<3$ with MNIST data.  Table \ref{tab:res_mnist_corrupt} shows accuracies on the validation set for the MNIST Corruption data for $n=4$. Table \ref{tab:frqi_mnist_multi} shows the results for Multi-Class classification between 0, 1 and 2 digit images using the VQC. For this, we split the $ez$ range into 3 parts.

\begin{table}[!htbp]
\centering
\caption{\textbf{Validation Set Accuracy using the VQC and the AC on FRQI representation for MNIST data}}
\label{tab:res_mnist}
\begin{tabular}{@{}cccccc@{}}
\toprule
\textbf{\begin{tabular}[c]{@{}c@{}}Training Set\\ Size\end{tabular}} & \textbf{Classifier} & \multicolumn{4}{c}{\textbf{$n$}} \\ \midrule
\textbf{} & \textbf{} & \textit{\textbf{1}} & \textit{\textbf{2}} & \textit{\textbf{3}} & \textit{\textbf{4}} \\\midrule
\multirow{2}{*}{\textbf{100}} & \textit{\textbf{VQC}} & 0.946 & 0.960 & 0.992 & 0.993 \\
 & \textit{\textbf{AC}} & 0.922 & 0.905 & 0.917 & 0.811 \\\midrule
\multirow{2}{*}{\textbf{200}} & \textit{\textbf{VQC}} & 0.938 & 0.960 & 0.995 & 0.996 \\
 & \textit{\textbf{AC}} & 0.904 & 0.877 & 0.793 & 0.691 \\\midrule
\multirow{2}{*}{\textbf{500}} & \textit{\textbf{VQC}} & 0.933 & 0.955 & 0.993 & 0.995 \\
 & \textit{\textbf{AC}} & 0.904 & 0.950 & 0.925 & 0.826 \\ \bottomrule
\end{tabular}
\end{table}

\begin{table}[!htbp]
\centering
\caption{\textbf{Validation Set Accuracy using the VQC and the AC on FRQI representation for MNIST Corruption data}}
\label{tab:res_mnist_corrupt}
\resizebox{\columnwidth}{!}{%
\begin{tabular}{@{}ccccccc@{}}
\toprule
\textbf{\begin{tabular}[c]{@{}c@{}}Training Set\\ Size\end{tabular}} & \textbf{Classifier} & \multicolumn{4}{c}{\textbf{Corruption}} &  \\ \midrule
 &  & \textit{\textbf{\begin{tabular}[c]{@{}c@{}}Shot\\ Noise\end{tabular}}} & \textit{\textbf{\begin{tabular}[c]{@{}c@{}}Impulse\\ Noise\end{tabular}}} & \textit{\textbf{\begin{tabular}[c]{@{}c@{}}Glass\\ Blur\end{tabular}}} & \textit{\textbf{\begin{tabular}[c]{@{}c@{}}Motion\\ Blur\end{tabular}}} & \textit{\textbf{Shear}} \\
\textbf{500} & \textit{\textbf{VQC}} & 0.996 & 0.998 & 0.994 & 0.990 & 0.997 \\
\textbf{} & \textit{\textbf{AC}} & 0.918 & 0.929 & 0.980 & 0.918 & 0.858 \\\midrule
 &  & \textit{\textbf{Scale}} & \textit{\textbf{Rotate}} & \textit{\textbf{Brightness}} & \textit{\textbf{Translate}} & \textit{\textbf{Stripe}} \\
\textbf{500} & \textit{\textbf{VQC}} & 0.988 & 0.993 & 0.993 & 0.965 & 0.993 \\
\textbf{} & \textit{\textbf{AC}} & 0.982 & 0.960 & 0.977 & 0.880 & 0.973 \\\midrule
 &  & \textit{\textbf{Fog}} & \textit{\textbf{Spatter}} & \textit{\textbf{\begin{tabular}[c]{@{}c@{}}Dotted\\ Line\end{tabular}}} & \textit{\textbf{ZigZag}} & \textit{\textbf{\begin{tabular}[c]{@{}c@{}}Canny\\ Edges\end{tabular}}} \\
\textbf{500} & \textit{\textbf{VQC}} & 0.988 & 0.995 & 0.997 & 0.992 & 0.989 \\
\textbf{} & \textit{\textbf{AC}} & 0.561 & 0.904 & 0.956 & 0.955 & 0.481 \\ \bottomrule
\end{tabular}%
}
\end{table}

\begin{table}[!htbp]
\centering
\caption{\textbf{Mulit-Class Validation Set Accuracy using the VQC on FRQI representation for MNIST data}}
\label{tab:frqi_mnist_multi}
\begin{tabular}{@{}ccccc@{}}
\toprule
\textbf{\begin{tabular}[c]{@{}c@{}}Training Set\\ Size\end{tabular}} & \multicolumn{4}{c}{\textbf{$n$}}                                                        \\ \midrule
                                                                        & \textit{\textbf{1}} & \textit{\textbf{2}} & \textit{\textbf{3}} & \textit{\textbf{4}} \\
\textbf{100} & 0.588 & 0.838  & 0.798 & 0.824 \\
\textbf{200} & 0.597 & 0.857  & 0.831 & 0.860 \\
\textbf{500} & 0.603 & 0.807 & 0.773 & 0.765 \\ \bottomrule
\end{tabular}
\end{table}

\subsection{\texorpdfstring{$2\times2$}{2 x 2} Color Images}

\begin{figure}[!htpb]
    \centering
    \includegraphics[width=\columnwidth]{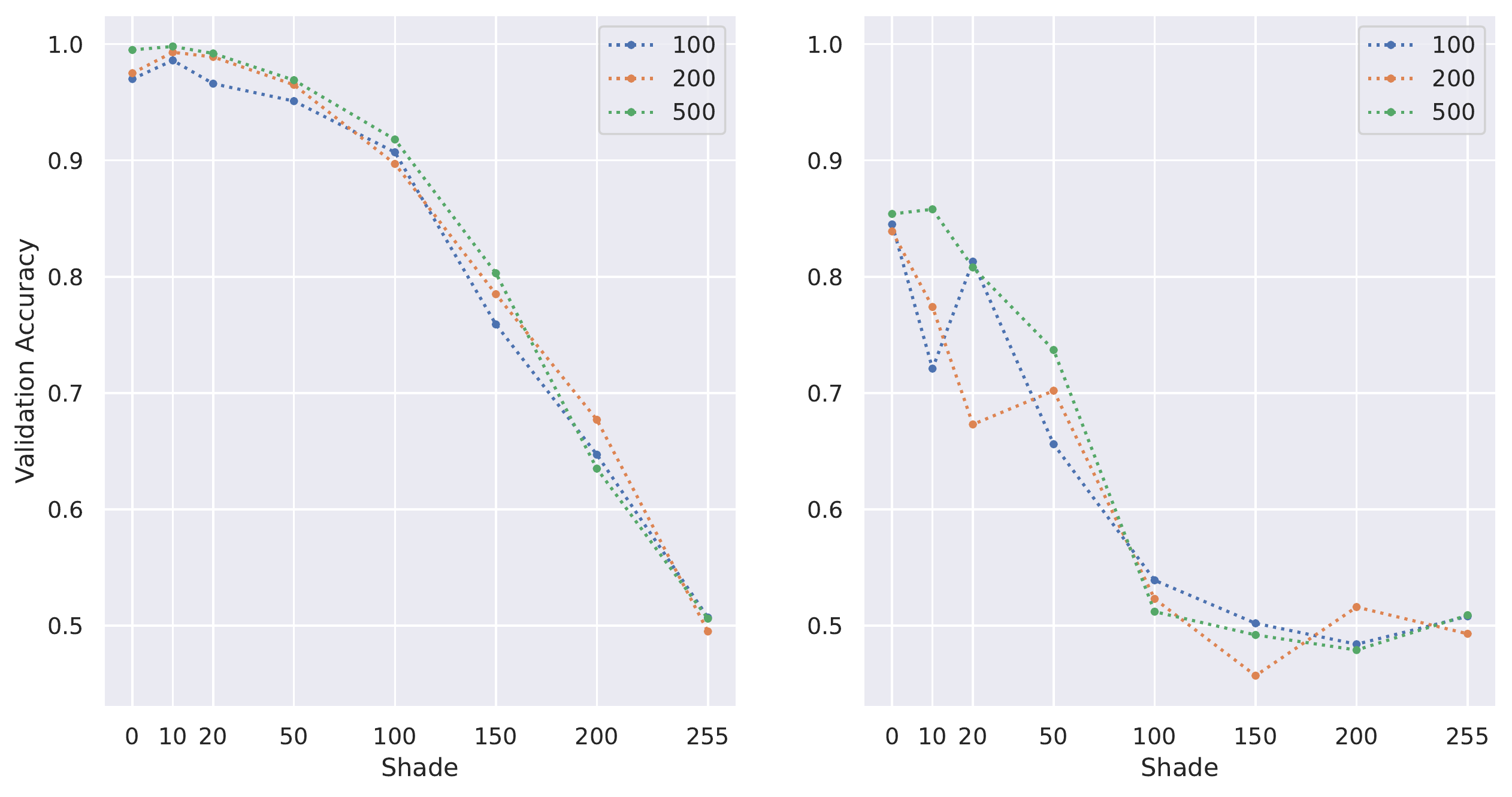}
    \caption{Validation accuracies as a function of shade with VQC on the left and AC on the right.}
    \label{fig:shades}
\end{figure}

Table \ref{tab:22_color} shows accuracy on the validation set using different training dataset sizes and shade values for the $2\times2$ colour image data. As expected the accuracy decreases as we increase the shade (see Fig \ref{fig:shades}). This is because, higher shade values imply smaller difference between the classes with a value of 255 implying no difference between the two classes.
\begin{table}[!htbp]
\centering
\caption{\textbf{Validation Set Accuracy using the VQC and the AC on MCQI representation for $2\times2$ Color Image data}}
\label{tab:22_color}
\begin{tabular}{@{}ccccccc@{}}
\toprule
\textbf{Shade} & \multicolumn{6}{c}{\textbf{Training Set Size}} \\ \midrule
\textbf{} & \multicolumn{2}{c}{\textit{\textbf{100}}} & \multicolumn{2}{c}{\textit{\textbf{200}}} & \multicolumn{2}{c}{\textit{\textbf{500}}} \\
\textbf{} & \textit{\textbf{VQC}} & \textit{\textbf{AC}} & \textit{\textbf{VQC}} & \textit{\textbf{AC}} & \textit{\textbf{VQC}} & \textit{\textbf{AC}} \\
\textbf{0} & 0.970 & 0.845 & 0.975 & 0.839 & 0.995 & 0.854 \\
\textbf{10} & 0.986 & 0.721 & 0.993 & 0.774 & 0.998 & 0.858 \\
\textbf{20} & 0.966 & 0.813 & 0.989 & 0.673 & 0.992 & 0.808 \\
\textbf{50} & 0.951 & 0.656 & 0.965 & 0.702 & 0.969 & 0.737 \\
\textbf{100} & 0.907 & 0.539 & 0.897 & 0.523 & 0.918 & 0.512 \\
\textbf{150} & 0.759 & 0.502 & 0.785 & 0.457 & 0.803 & 0.492 \\
\textbf{200} & 0.647 & 0.484 & 0.677 & 0.516 & 0.635 & 0.479 \\
\textbf{255} & 0.507 & 0.508 & 0.495 & 0.493 & 0.506 & 0.509 \\ \bottomrule
\end{tabular}
\end{table}
\section{CONCLUSIONS AND FUTURE WORK}
\label{conclude}
Encouraging results on benchmark datasets have been obtained with both VQC and AC for binary and multi-class image classification. The work can be expanded to classify more involved images. The ansatz used for VQC was a simple layer. More research can be done to find a better ansatz that can provide improved performance while using a lower number of epochs. The effect of noise and shots on the performance can also be studied. Similarly, different autoencoder models, for example, denoising autoencoder, can be studied. Also, one can look into other Image Processing tasks like filtering on the representations/encodings.

\balance
\printbibliography

\end{document}